\documentclass[10pt,fleqn,a4paper]{article}

\pdfoutput=1

\usepackage[euler-digits]{eulervm}

\usepackage{amsmath}
\usepackage{amssymb}
\usepackage{graphicx}
\usepackage[numbers]{natbib}
\usepackage{titlesec}
\usepackage{amsmath}
\usepackage{dsfont}
\usepackage{amsthm}
\usepackage{multicol}
\usepackage{bold-extra}
\usepackage[title]{appendix}
\usepackage[usenames,dvipsnames,svgnames,table]{xcolor}
\usepackage[margin=30pt,font=small,labelfont=bf,labelsep=endash]{caption}
\usepackage{amsthm}
\usepackage{rotating}
\usepackage{hyperref}
\usepackage{slashed} 
\usepackage{epigraph}
\usepackage{empheq}

\newcommand{\trans}[1]{{#1}^{\scriptscriptstyle{\rm T}}}

\newtheorem{mydef}{Definition}
\newtheorem{myprop}{Proposition}
\newtheorem{myremark}{Remark}
\numberwithin{equation}{section}
 
\definecolor{myblue}{rgb}{.9, .9, 0.9}
\newcommand*\mybluebox[1]{%
\colorbox{myblue}{\hspace{1em}#1\hspace{1em}}}

\begin{document}
\title{Use of Dirichlet Distributions and Orthogonal Projection
  Techniques for the Fluctuation Analysis of Steady--State
  \\ Multivariate Birth--Death Systems}
\author{
\\[-0.1cm]
{{{Filippo Palombi$^{a}$\footnote{Corresponding
        author. E--mail: {\tt filippo.palombi@enea.it}}\ \  and Simona Toti$^{b}$}}}\\[1.0ex]
 {{\small{$^a$ENEA -- Italian Agency for New Technologies, Energy and}}}\\
{{\small{Sustainable Economic Development,}}}
 {\small {{\it Via Enrico Fermi 45, 00040 Frascati -- Italy}}}\\[.1cm]
 {{\small{$^b$ISTAT -- Istituto Nazionale di Statistica,}}}
 {\small {\it Via Cesare Balbo 16, 00184 Rome -- Italy}}\\[.1cm]
}

\date{July 2014}

\newcommand{\cA}{{\cal A}}
\newcommand{\cB}{{\cal B}}
\newcommand{\cC}{{\cal C}}
\newcommand{\cD}{{\cal D}}
\newcommand{\cE}{{\cal E}}
\newcommand{\cF}{{\cal F}}
\newcommand{\cG}{{\cal G}}
\newcommand{\cH}{{\cal H}}
\newcommand{\cI}{{\cal I}}
\newcommand{\cJ}{{\cal J}}
\newcommand{\cK}{{\cal K}}
\newcommand{\cL}{{\cal L}}
\newcommand{\cM}{{\cal M}}
\newcommand{\cN}{{\cal N}}
\newcommand{\cO}{{\cal O}}
\newcommand{\cP}{{\cal P}}
\newcommand{\cQ}{{\cal Q}}
\newcommand{\cR}{{\cal R}}
\newcommand{\cS}{{\cal S}}
\newcommand{\cT}{{\cal T}}
\newcommand{\cU}{{\cal U}}
\newcommand{\cV}{{\cal V}}
\newcommand{\cW}{{\cal W}}
\newcommand{\cX}{{\cal X}}
\newcommand{\cY}{{\cal Y}}
\newcommand{\cZ}{{\cal Z}}
\newcommand{\indicator}[1]{{\mathds{I}\{#1\}}}
\newcommand{\dC}{\mathds{C}}
\newcommand{\dE}{\mathds{E}}
\newcommand{\dP}{\mathds{P}}
\newcommand{\dR}{\mathds{R}}
\newcommand{\dN}{\mathds{N}}
\newcommand{\dK}{\mathds{K}}
\newcommand{\dZ}{\mathds{Z}}
\newcommand{\rd}{\text{d}}
\newcommand{\re}{\text{e}}
\newcommand{\rO}{\text{O}}
\newcommand{\rr}{\text{r}}
\newcommand{\ri}{\text{i}}
\newcommand{\rx}{\text{x}}
\newcommand{\sh}{\text{sh}}
\newcommand{\eff}{\text{eff}}
\newcommand{\diag}{\text{diag}}
\newcommand{\off}{\text{off--diag}}
\newcommand{\ml}{m\phantom{l}}
\newcommand{\mi}{\mathrm{i}} 
\newcommand{\me}{\mathrm{e}} 
\newcommand{\fA}{{\mathfrak A}}
\newcommand{\fB}{{\mathfrak B}}
\newcommand{\fC}{{\mathfrak C}}
\newcommand{\fD}{{\mathfrak D}}
\newcommand{\fE}{{\mathfrak E}}
\newcommand{\fF}{{\mathfrak F}}
\newcommand{\fG}{{\mathfrak G}}
\newcommand{\fH}{{\mathfrak H}}
\newcommand{\fI}{{\mathfrak I}}
\newcommand{\fJ}{{\mathfrak J}}
\newcommand{\fK}{{\mathfrak K}}
\newcommand{\fL}{{\mathfrak L}}
\newcommand{\fM}{{\mathfrak M}}
\newcommand{\fN}{{\mathfrak N}}
\newcommand{\fO}{{\mathfrak O}}
\newcommand{\fP}{{\mathfrak P}}
\newcommand{\fQ}{{\mathfrak Q}}
\newcommand{\fR}{{\mathfrak R}}
\newcommand{\fS}{{\mathfrak S}}
\newcommand{\fT}{{\mathfrak T}}
\newcommand{\fU}{{\mathfrak U}}
\newcommand{\fV}{{\mathfrak V}}
\newcommand{\fW}{{\mathfrak W}}
\newcommand{\fX}{{\mathfrak X}}
\newcommand{\fY}{{\mathfrak Y}}
\newcommand{\fZ}{{\mathfrak Z}}

\newcommand{\aka}{{\it a.k.a.}}
\newcommand{\etc}{{\it etc}}
\newcommand{\ie}{{\it i.e.\ }}
\newcommand{\eg}{{\it e.g.\ }}
\newcommand{\cfr}{{\it cf.\ }}
\newcommand{\adj}[1]{\text{adj}_V(#1)}
\newcommand{\gen}{{\Omega_{\text{vt}}}}
\newcommand{\St}{{S_{\text{vt}}}}
\newcommand{\genp}{{\Omega}}
\newcommand{\Erdos}{{Erd\H{o}s}}
\newcommand{\dPer}{\mathds{P}_{\rm er}}
\newcommand{\E}{\mathds{E}}
\newcommand{\kI}{{\kappa_{\scriptscriptstyle\text{iso}}}}
\newcommand{\cLFP}{{\cal L}_{\scriptscriptstyle{\rm FP}}}
\newcommand{\nbc}{{n}_{\scriptscriptstyle{\rm bc}}}

\maketitle
\begin{abstract}
Approximate weak solutions of the Fokker--Planck equation 
represent a useful tool to analyze the equilibrium fluctuations
of birth--death systems, as they provide a quantitative
knowledge lying in between numerical simulations and exact analytic
arguments. In the  present paper, we adapt the general mathematical
formalism known as the Ritz--Galerkin method for partial differential
equations to the Fokker--Planck equation with
time--independent polynomial drift and diffusion coefficients on the
simplex. Then, we show how the method works in two examples, namely
the binary and multi--state voter models with zealots.
\end{abstract}

\section{Introduction}

Multivariate birth--death models have since long captured the interest
of researchers in statistical physics as they represent a natural
mathematical framework to investigate a plethora of interdisciplinary
problems, ranging from opinion diffusion to language emergence, cultural
dissemination and epidemic
spreading~\cite{FortCastLor,Keeling}. Broadly speaking, such    
models describe an evolving population of agents, each lying in one of
$Q$ allowed physical states. The system is macroscopically represented by a
state vector $\phi=(\phi_k)_{k=1}^Q$, with $\phi_k$ denoting the fraction of
agents in the $k$--th state. By definition, for
$Q<\infty$ the state vector lives on the $Q$--simplex
\begin{equation}
  T_Q(s) = \left\{\phi\in\dR_+^Q:\quad \sum_{k=1}^Q\phi_k = s\right\}\,.
  \label{eq:simplex}
\end{equation}
The meaning of the parameter $s$ will become clear in the sequel,
while for the time being the reader may assume $s=1$. If the
microscopic dynamics of the model is determined by Markovian agent--agent
interactions altering the components of $\phi$, then in the
thermodynamic limit the system is known to obey a Fokker--Planck equation (FPE)
(see for instance ref. \cite[chapt.~7]{Gardiner}),   
\begin{align}
  & \label{eq:fpe}\partial_t \cP(t,\bar\phi) = -\sum_{k=1}^{Q-1}
  \partial_k\left[A_k(t,\bar\phi)\cP(t,\bar\phi)\right] +
  \frac{1}{2}\sum_{i,k=1}^{Q-1}\partial_i\partial_k\left[B_{ik}(t,\bar\phi)
    \cP(t,\bar\phi)\right] \equiv\cLFP\cdot\cP(t,\bar\phi)\,,\\[1.0ex]
  & \label{eq:bc} \cP(t,\bar\phi) = 0 \quad \text{if}\quad \phi\notin T_Q(s)\,.
\end{align}
where $\bar\phi=(\phi_k)_{k=1}^{Q-1}$ denotes the essential state vector
 (obtained from $\phi$ by conventionally leaving out its $Q$--th
component), $\cP(t,\bar\phi)$ represents the probability density of 
$\bar\phi$ at time $t$ and we define $\partial_t \equiv
\partial/\partial t$ and $\partial_k \equiv
\partial/\partial\phi_k$. $\cLFP$ is commonly referred to as
the Fokker--Planck operator. We do not impose any initial condition to
eq.~(\ref{eq:fpe}), as this  is not relevant to our aims.   

We shall make the assumption -- valid for several birth--death models
-- that the drift coefficients $(A_k)_{k=1}^{Q-1}$ and the diffusion
ones  $(B_{ik})_{i,k=1}^{Q-1}$ are time--independent polynomials of the
components of $\bar\phi$. We shall also assume that the stochastic
dynamics of the model has no exit states such as consensus or
no--infected--agents configurations. If this the case, the system is
expected to asymptotically relax to a dynamic equilibrium, with
$\phi$ wandering across $T_Q(s)$ according to the stochastic process  
\begin{equation}
  \rd\phi_\ell(t) = A_\ell(\bar\phi)\rd t + \sum_{k=1}^{Q-1}C_{\ell
    k}(\bar\phi)\rd W_k(t) + \rd K_\ell(t)\,,\qquad
  \ell=1,\ldots,Q-1\,, 
  \label{eq:stochproc}
\end{equation}
and eventually distributing according to a limit probability density
$\cP(\bar\phi)=\lim_{t\to\infty}\cP(t,\bar\phi)$. Here, the matrix $C$
is related to the diffusion matrix  via $B = C \cdot \trans{C}$
($\trans{C}$ is the transposed matrix of $C$), $W(t)=\left(W_k(t)\right)_{k=1}^{Q-1}$ is a
Wiener process describing the stochastic diffusion of the state vector
and $K(t) = \left(K_k(t)\right)_{k=1}^{Q-1}$ is a Skorokhod bounded variation
process~\cite{Skorokhod}, increasing only when $\phi\in\partial
T_Q(s)$ so as to ensure the boundary condition, eq.~(\ref{eq:bc}). 

Numerical simulations of eq.~(\ref{eq:stochproc}) can be efficiently
used to make quantitative statements on $\cP(\bar\phi)$, yet they give
little insight on its analytic structure. From this point of view, a
more convenient approach would be to represent the equilibrium
distribution in terms of a properly chosen function basis. A
legitimate possibility is to consider polynomial distributions on the
$Q$--simplex. In regard to this choice, we recall that a Dirichlet
distribution 
$\text{Dir}(\gamma)$ of order $Q$ with parameter $\gamma\in\dN^Q$, has
probability density  
\begin{align}
  \label{eq:Dirichlet}
  & \cD_\gamma(\bar\phi) =
  \frac{\Gamma(|\gamma|)}{\prod_{k=1}^{Q}\Gamma(\gamma_k)}s^{1-|\gamma|}\left(
  \prod_{k=1}^{Q-1}
  \phi_k^{\slashed{\gamma}_k}\right)(s-|\bar\phi|)^{\slashed{\gamma}_Q}\,,\qquad
  \bar\phi\in\bar T_{Q}(s)\,,\\[1.0ex] 
  & \bar T_Q(s) = \{\bar\phi\in\dR_+^{Q-1}: \ |\bar\phi|\le s\}\,,
\end{align}
with $\slashed{\gamma}_k \equiv \gamma_k-1$ and $|x| \equiv \sum
x_k$. We notice that $\bar T_Q(s)$ is in one--to--one correspondence
with $T_Q(s)$, hence we can equivalently write $\phi\in T_Q(s)$ or 
$\bar\phi\in\bar T_Q(s)$. It is crucial for the reader who is
unfamiliar with the mathematics of the simplex to learn how to
calculate Dirichlet integrals, \ie polynomial integrals on $\bar
T_Q(s)$. As an example, we review in App.~A an elegant way to work out
the normalization constant of eq.~(\ref{eq:Dirichlet}). This is
sufficient to be able to reproduce all calculations presented in the
paper. That being said, an important feature of the Dirichlet
distributions is represented by 
\begin{myprop} Dirichlet distributions with positive integer
  indices provide a basis of polynomials, that is to say
  \vskip -0.5cm
  \begin{equation}
    {\rm span}\left\{\cD_\gamma(\bar\phi):\ \gamma\in\dN^Q \text{ and
    } \ |\slashed{\gamma}| = 
    n\right\} = {\rm
      span}\left\{\bar\phi^\alpha:\ \alpha\in\dN_0^{Q-1} \text{  and }
    |\alpha|\le n\right\}\,, 
    \label{eq:span}
  \end{equation}
  where we make use of the multi--index notation $\bar\phi^\alpha
  \equiv \phi_1^{\alpha_1}\cdot\ldots\cdot\phi_{Q-1}^{\alpha_{Q-1}}$.  
\end{myprop}
\noindent\emph{Proof.} Given $\gamma\in\dN^{Q}$ with
$|\slashed{\gamma}| = n$, $\cD_\gamma(\bar\phi)$ is a polynomial with 
degree~$n$, hence it can be written as a linear combination of
monomials with degree $\le n$. Conversely, suppose that $\alpha\in
\dN_0^{Q-1}$ and $|\alpha|=n$. Then $\bar\phi^\alpha \propto
\cD_{\gamma}(\bar\phi)$ with $\gamma =
(\alpha_1+1,\ldots,\alpha_{Q-1}+1,1)$. Finally, if $\alpha\in
\dN_0^{Q-1}$ and $|\alpha|=m<n$, then we define $\gamma =
(\alpha_1+1,\ldots,\alpha_{Q-1}+1,n-m+1)$ such that
$|\slashed{\gamma}|=n$, and we observe that 
\begin{equation}
  \cD_{\gamma}(\bar\phi) \propto \bar\phi^{\alpha}(s-|\bar\phi|)^{n-m}
  \propto \bar\phi^\alpha + \cE_\alpha(\bar\phi) 
\end{equation}
with $\cE_\alpha(\bar\phi)$ being a linear combination of monomials,
each with degree $>m$. Therefore, the proof can be obtained by
backward induction on $m=n-1,n-2,\ldots.$\qed 

\noindent  Motivated by this observation, we introduce a polynomial
approximation $\cP_n$ to $\cP$ with degree $n$, reading 
\begin{align}
  \label{eq:polexp}
  \cP_n(\bar\phi) & = \sum_{\gamma\in\Omega_n}c_\gamma
  \cD_\gamma(\bar\phi)\,,\qquad \Omega_n =
  \{\gamma\in\dN^Q:\ |\slashed{\gamma}|=n\}\,,\\[1.0ex] 
  \label{eq:polnorm}
  \sum_{\gamma\in\Omega_n}c_\gamma & = 1\,.
\end{align}
Owing to Prop.~1, eq.~(\ref{eq:polexp}) is equivalent to a complete
sum over all monomials of degree $\le n$, while eq.~(\ref{eq:polnorm})
is just obtained by imposing that $\cP_n$ is correctly normalized on
$\bar T_Q(s)$. Sometimes, $\Omega_n$ is referred to by mathematicians
as the \emph{bucket space}. The choice of the Dirichlet distributions
as a polynomial basis is favourable for several reasons, as we shall 
explain in next sections.  

The aim of the present paper is to describe how estimates of the
expansion coefficients $c \equiv \{c_\alpha\}$ can be determined
straightaway from the FPE, with a view to providing a hopefully
helpful analysis tool to practitioners in the physics of complex
systems. To this end, we adapt to eqs.~(\ref{eq:fpe})--(\ref{eq:bc}) a
mathematical technique known as the Ritz--Galerkin (RG) method for
partial differential equations (see for instance~\cite{finitel} for
a technical introduction), which is commonly used by engineers in many
applicative fields, including fluid and solid mechanics,
hydrodynamics, wave propagation, electromagnetism and many
others~\cite{spectral}. Our approach makes use of orthogonal
polynomials on $\bar T_Q(s)$ as test functions and point--like
zero--orthogonal--flux conditions on $\partial \bar T_Q(s)$.   

The paper, which is written in a pedagogical style with detailed
calculations, is organized as follows. In sect.~2, we provide a short
compendium of orthogonal polynomials on $\bar T_Q(s)$, while in
sect.~3 we review the basics of the RG method and discuss
how to apply it to the FPE for birth--death models with polynomial
drift and diffusion coefficients. In sects.~4 and~5, we show
applications of the method respectively to the binary voter model with
zealots studied in \cite{Mobilia2007}, for which an exact solution of
the FPE is known, and to its generalization to the multi--state
case. In sect.~6, we discuss how symmetry arguments can help reduce
the computational budget needed to implement the method. We finally
draw our conclusions in sect.~7.

\section{Orthogonal polynomials on the simplex}

Since $\cD_\gamma$ is a polynomial distribution on $\bar
T_Q(s)$ with degree $n$ for $\gamma\in\Omega_n$, it is rather natural
to look for orthogonal polynomial bases on the simplex. The
general theory of multivariate orthogonal polynomials is still an open
research field: it does not belong to the average undergraduate
background of physicists and is not even discussed in many essays in
the mathematical literature. Fortunately, an excellent introduction is
provided in ref.~\cite{XuBook}. We refer the reader to that book for a
comprehensive presentation of classical and recent developments on the
subject, while for the sake of readability and self--consistency of
the paper we review here those aspects which are closely related to
our ends.  

\newpage

First of all, a multivariate polynomial $P_\alpha$ on $\bar T_Q(s)$,
indexed by $\alpha\in\dN_0^{Q-1}$ can be always represented by its
monomial expansion  
\begin{equation}
P_\alpha(\bar\phi) = \sum_{\beta\le\alpha} c_\beta \bar\phi^\beta
\equiv \sum_{\beta_1=0}^{\alpha_1}\ldots
\sum_{\beta_{Q-1}=0}^{\alpha_{Q-1}} 
c_{\beta_1\ldots\beta_{Q-1}}\phi_1^{\beta_1}\ldots
\phi_{Q-1}^{\beta_{Q-1}}\,. 
\end{equation}
The degree of $P_\alpha$ is the maximum degree of its monomials, \ie
$\deg\{P_\alpha\}=|\alpha|$. Secondly, the orthogonality notion on
$\bar T_Q(s)$ depends on the introduction of a scalar product, which
in turn requires the specification of a measure. The standard choice
-- which we adopt here -- is to weight the Lebesgue measure by a
Dirichlet distribution, \ie to define 
\begin{equation}
\langle f,g\rangle_\kappa = \int_{\bar T_Q(s)}\rd
\bar\phi\ f(\bar\phi)g(\bar\phi)\cD_\kappa(\bar\phi)\,,\qquad
\kappa\in\dN^{Q}\,, 
\label{eq:defscalprod}
\end{equation}
for sufficiently regular functions $f,g$ on $\bar T_Q(s)$. The
Dirichlet weight is such that $\langle 1,1\rangle_\kappa =
1$. Thirdly, two polynomials $P$ and $Q$ on $\bar T_Q(s)$ are said to
be orthogonal if $\langle P,Q\rangle_\kappa = 0$, while a polynomial
$P$ is called an orthogonal polynomial if it is orthogonal to all
polynomials of lower degree, \ie 
\begin{equation}
\langle P,Q\rangle_\kappa = 0\,, \qquad \forall Q \quad \text{ with
}\quad  \deg\{Q\}<\deg\{P\}\,. 
\end{equation}
The main difference between orthogonal polynomial bases in one and
several variables, is that the former count just one element per
degree, whereas the latter count many of them. To be precise, it can
be shown as a trivial consequence of Prop.~1 that 
\begin{equation}
\dim\left\{\text{orthogonal polynomials } P \text{ on } \bar
T_Q(s):\ \deg\{P\}\le n \right\} = |\Omega_n| = {n+Q-1 \choose n}\,. 
\end{equation}
Notice that the \emph{bucket space} expands roughly as 
\begin{equation}
|\Omega_n| \approx
\frac{\exp\left\{(Q-1)(H_n-\gamma_{\scriptscriptstyle{\rm
      E}})\right\}}{\Gamma(Q)}\,,\qquad H_n = \sum_{k=1}^n
\frac{1}{k}\,, 
\end{equation}
with $H_n$ being the $n$--th harmonic number and
$\gamma_{\scriptscriptstyle{\rm E}} = 0.57721...$  the
Euler--Mascheroni constant. For this reason, the RG method becomes
computationally challenging even for models with a moderately large
value of $Q$.   

Now, there exist several sets of orthogonal polynomials
with respect to eq.~(\ref{eq:defscalprod}). Along with
ref.~\cite{XuBook}, we focus on two of them, namely
\vskip 0.2cm
{\raggedleft \underline{the monomial basis}}
\begin{equation}
\label{eq:Vbasis}
V_\alpha(\bar\phi) = \sum_{\beta\le\alpha}
(-1)^{|\alpha|+|\beta|}s^{-|\beta|}\prod_{i=1}^{Q-1}{\alpha_i \choose
  \beta_i}\dfrac{({\kappa_i})_{\alpha_i}}{({\kappa_i})_{\beta_i}}
\dfrac{(|\kappa|-1)_{|\alpha|+|\beta|}}{(|\kappa|-1)_{2|\alpha|}}\,\bar\phi^\beta  
\equiv  \sum_{\beta\le\alpha}v_{\alpha\beta}(\kappa)\bar\phi^\beta\,, 
\end{equation}
{\raggedleft \underline{the Appel basis}}
\begin{equation}
\label{eq:Ubasis}
U_\alpha(\bar\phi) =
\cD_\kappa(\bar\phi)^{-1}\partial^{|\alpha|}_\alpha
\left[\phi_1^{\alpha_1+\kappa_1-1}\ldots 
  \phi_{Q-1}^{\alpha_{Q-1}+\kappa_{Q-1}-1}
  (s-|\bar\phi|)^{|\alpha|+\kappa_{Q}-1} \right]\,, 
\end{equation}
for $\alpha\in\dN_0^{Q-1}$, with $\partial^{|\alpha|}_\alpha \equiv
\partial^{|\alpha|}/\partial x_1^{\alpha_1}\ldots \partial
x_d^{\alpha_d}$ and with $(x)_n = x(x+1)\ldots (x+n-1)$ denoting the
Pochhammer symbol (also known as the raising factorial). The
polynomials $\{V_\alpha\}$ and $\{U_\alpha\}$ fulfill the following
properties: 
\begin{myprop}
For any polynomial $P_\beta$ on $\bar T_Q(s)$, it holds
\begin{equation}
\langle V_\alpha,P_\beta\rangle_\kappa = \langle
U_\alpha,P_\beta,\rangle_\kappa = 0 \quad \text{if} \quad 
  |\beta| < |\alpha|\,.
\end{equation}
Moreover, the polynomials $\{V_\alpha:\,\alpha\in\dN_0^{Q-1}\}$ and
$\{U_\alpha:\,\alpha\in\dN_0^{Q-1}\}$ are biorthogonal, \ie they
fulfill 
\begin{equation}
\langle V_\alpha, U_\beta\rangle_\kappa = f_\alpha\,
\delta_{\alpha\beta}\,,\qquad f_\alpha = s^{|\alpha|} 
  \dfrac{\left[\prod_{m=1}^{Q-1}
    (\kappa_m)_{\alpha_m}\Gamma(\alpha_m+1)\right](\kappa_{Q})_{|\alpha|}}
        {(|\kappa|)_{2|\alpha|}}\,.  
\end{equation}
\end{myprop}
\noindent\emph{Proof.} The proof is contained in
ref.~\cite[chap.~2]{XuBook}. Here, we only review the argument showing
that $\langle V_\alpha, P_\beta\rangle_\kappa=0$, since we shall need
a formula which is derived along the proof. Owing to Prop. 1, it is
sufficient to prove that $\langle V_\alpha,X_\gamma
\rangle_{\kappa}=0$ for 
\begin{equation}
X_\gamma(\bar\phi) =
\,\left[\prod_{m=1}^{Q-1}\phi_m^{\slashed{\gamma}_m}\right]
(s-|\bar\phi|)^{\slashed{\gamma}_{Q}}\,,   
\end{equation}
with $\gamma\in\dN^{Q}$ and $|\slashed{\gamma}|=|\alpha|-1$. Indeed, it holds
\begin{align}
\langle V_\alpha,X_\gamma\rangle_\kappa & = \int_{\bar T_Q(s)}\rd\bar\phi\ 
V_\alpha(\bar\phi) X_\gamma(\bar\phi)  \cD_{\kappa}(\bar\phi)\nonumber\\[1.0ex]
& =
\frac{\Gamma(|\kappa|)}{\prod_{k=1}^{Q}\Gamma(\kappa_k)}s^{1-|\kappa|}
\sum_{\beta\le\alpha}(-1)^{|\alpha|+|\beta|}s^{-|\beta|}
\prod_{m=1}^{Q-1}{\alpha_m \choose
  \beta_m}\dfrac{({\kappa_m})_{\alpha_m}} 
{({\kappa_m})_{\beta_m}}\dfrac{(|\kappa|-1)_{|\alpha|+|\beta|}}
{(|\kappa|-1)_{2|\alpha|}}\nonumber\\[0.0ex]
\phantom{\langle V_\alpha,X_\gamma\rangle_\kappa} & \cdot \int_{\bar
  T_Q(s)}\rd\bar\phi\  \prod_{\ell=1}^{Q-1}\phi_\ell^{\beta_\ell+
  \slashed{\gamma}_\ell+\kappa_\ell-1}
(s-|\bar\phi|)^{\kappa_{Q}+\slashed{\gamma}_{Q}-1}\nonumber\\[1.0ex]  
& =
s^{|\slashed{\gamma}|}\frac{\Gamma(|\kappa|)}{\Gamma(|\kappa|+2|\alpha|-1)}
\left[\prod_{m=1}^{Q-1}\frac{\Gamma(\kappa_m+\alpha_m)\Gamma(\slashed{\gamma}_m+1)}
  {\Gamma(\kappa_m)}\right]\frac{\Gamma(\kappa_{Q}+\slashed{\gamma}_{Q})}
     {\Gamma(\kappa_Q)}\nonumber\\[0.0ex] 
& \cdot
     \sum_{\beta\le\alpha}\prod_{m=1}^{Q-1}(-1)^{\beta_m}{\alpha_m\choose\beta_m}
         {\kappa_m+\slashed{\gamma}_m+\beta_m-1\choose
           \slashed{\gamma}_m} \frac{\Gamma(|\kappa| + |\alpha| +
           |\beta|-1)} {\Gamma(|\kappa| + |\slashed{\gamma}| + |\beta|)}\,.
\end{align}
If $|\slashed{\gamma}|=|\alpha|-1$, the rightmost ratio of
$\Gamma$--functions simplifies and we are left with a product of
independent sums. From a Chu--Vandermonde formula 
\begin{equation}
\sum_{k=0}^n(-1)^k{n\choose k}{a+k\choose m} =
\frac{(-m)_n}{\Gamma(m+1)(1+a)_{n-m}}\,,\qquad
m,n\in\dN\,,\ a\in\dR\,, 
\end{equation}
it follows
\begin{align}
\langle V_\alpha,X_\gamma\rangle_\kappa & =
s^{|\slashed{\gamma}|}\frac{\Gamma(|\kappa|)}{\Gamma(|\kappa|+2|\alpha|-1)}
\left[\prod_{m=1}^{Q-1}\frac{\Gamma(\kappa_m+\alpha_m)\Gamma(\slashed{\gamma}_m+1)}
  {\Gamma(\kappa_m)}\right]\frac{\Gamma(\kappa_{Q}+
  \slashed{\gamma}_{Q})}{\Gamma(\kappa_Q)}\nonumber\\[0.0ex]  
& \cdot
\prod_{m=1}^{Q-1}\frac{(-\slashed{\gamma}_m)_{\alpha_m}}
     {\Gamma(\slashed{\gamma}_m+1)(\kappa_m+\slashed{\gamma}_m)_{\alpha_m-\slashed{\gamma}_m}}\,.  
\end{align}
However, we know that $(-m)_n=0$ for $m<n$. Since
$|\slashed{\gamma}|=|\alpha|-1$, there is at least one value of $m$
for which $\slashed{\gamma}_m<\alpha_m$. Therefore we conclude that
$V_\alpha$ is orthogonal to $X_\gamma$. \qed 

\begin{myremark}
The polynomials $\{V_\alpha\}$ can be easily coded. Indeed, we use
them for computations. However, it should be noticed that the
numerical evaluation of eq.~(\ref{eq:Vbasis}) can be critical,
specially for $|\alpha|\gg 1$, since $V_\alpha$ adds largely different
ratios of factorials with alternating signs. For this reason,
computations should be performed and crosschecked with different
levels of floating point rounding. Most of the numerical experiments
described in next sections have been done in Maple$\texttrademark$,
which allows to control the numerical precision by the environment
variable {\tt Digits}. We employ the biorthogonal basis $\{U_\alpha\}$
essentially to develop theoretical arguments.  \qed
\end{myremark}

\begin{figure}[t!]
    \centering
    \includegraphics[width=0.6\textwidth]{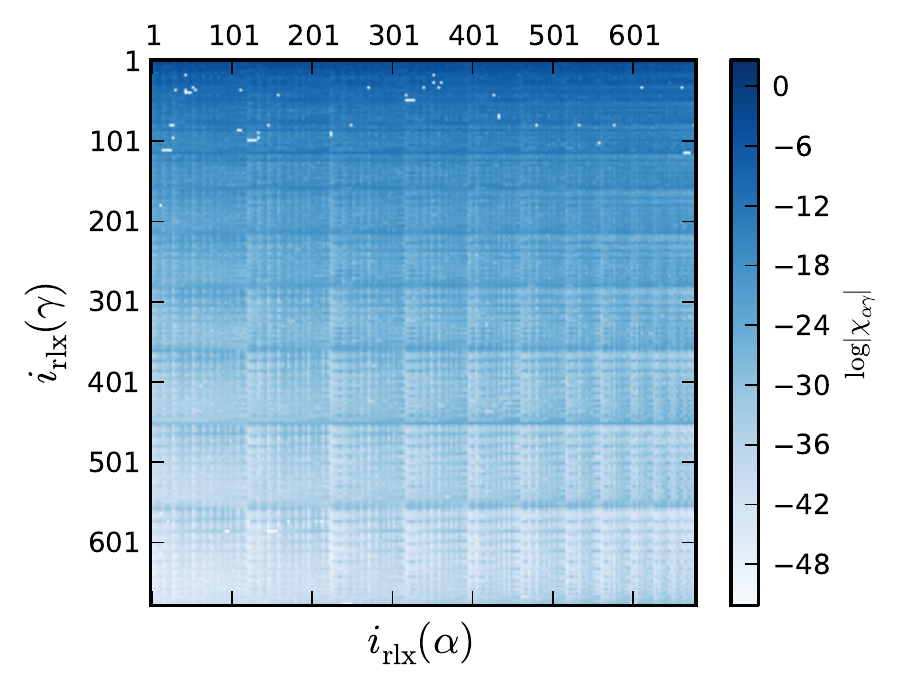}
  \caption{ \footnotesize Heat map of $\log|\chi_{\alpha\gamma}|$ for
    $Q=4$, $n=14$, $\kappa = (2,2,2,2)$ and $s=1$. In this case
    $|\Omega_n| = 680$.} 
  \label{fig:chi}
\end{figure}

\noindent If $|\slashed{\gamma}|\ne |\alpha|-1$, the factorization
property does not hold, hence we are left with the general formula 
\begin{empheq}[box=\mybluebox]{align}
\label{eq:scalprod}
\chi_{\alpha\gamma} \equiv \langle V_\alpha,\cD_\gamma\rangle_\kappa &
= s^{1-Q}
\frac{\Gamma(|\gamma|)\Gamma(|\kappa|)}{\Gamma(|\kappa|+2|\alpha|-1)}
\left[\prod_{m=1}^{Q-1}\frac{\Gamma(\kappa_m+\alpha_m)}{\Gamma(\kappa_m)}\right]\frac{\Gamma(\kappa_{Q}+{\gamma}_{Q}-1)}{\Gamma(\kappa_Q)\Gamma(\gamma_Q)}\nonumber\\[0.0ex] 
& \hskip -0.4cm \cdot \sum_{\beta\le\alpha}\frac{\Gamma(|\kappa| +
  |\alpha| + |\beta|-1)}{\Gamma(|\kappa| + |{\gamma}| +
  |\beta|-Q)}\prod_{m=1}^{Q-1}(-1)^{\beta_m}{\alpha_m\choose\beta_m}{\kappa_m+{\gamma}_m+\beta_m-2\choose
  {\gamma}_m-1}\,, 
\end{empheq}
valid for $|\alpha|\le|\slashed{\gamma}|$. We can extend Remark 1 to
eq.~(\ref{eq:scalprod}) as well. The meaning of the matrix $\chi =
\{\chi_{\alpha\gamma}\}$ becomes clear if we expand $\cD_\gamma$ along
the Appel basis, namely 
\begin{equation}
\cD_{\gamma}(\bar\phi) =
\sum_{0\le|\beta|\le|\slashed{\gamma}|}d_{\gamma\beta}\,U_\beta(\bar\phi)\,. 
\label{eq:Dirproj}
\end{equation}
By projecting both sides of eq.~(\ref{eq:Dirproj}) onto $V_\alpha$, we
obtain 
\begin{align}
\chi_{\alpha\gamma} & = \langle V_\alpha,\cD_{\gamma}\rangle_\kappa =
\sum_{0\le|\beta|\le|\slashed{\gamma}|}d_{\gamma\beta}\, \langle
V_\alpha,U_\beta\rangle_\kappa  = \sum_{0\le|\beta|\le
  |\slashed{\gamma}|}d_{\gamma\beta}\, f_\beta\, \delta_{\alpha\beta}
=  d_{\gamma\alpha}\,f_\alpha\,\,, 
\end{align}
whence it follows
\begin{equation}
\cD_\gamma(\bar\phi) = \sum_{0\le|\beta|\le|\slashed{\gamma}|}\frac{\chi_{\beta\gamma}}{f_\beta}\,U_\beta(\bar\phi)\,.
\end{equation}
We thus conclude that, given $n\ge 0$, $\gamma\in\Omega_n$ and
$\alpha\in\dN_0^{Q-1}$ such that $|\alpha|\le |\slashed{\gamma}|$, the
matrix elements $\chi_{\alpha\gamma}$ are essentially the expansion
coefficients of the  (non--orthogonal) basis $\{\cD_\gamma\}$ along
the (orthogonal) Appel basis $\{U_\alpha\}$, \ie $\chi$ is essentially
a change--of--basis matrix. It is interesting to look at the numerical 
values of $\chi_{\alpha\gamma}$ in some specific case. As an example,
in Fig.~1 we show a heat map of $\log|\chi_{\alpha\gamma}|$ for
$Q=4$, $n=14$, $\kappa = (2,2,2,2)$ and $s=1$; here, the index
arrays are sorted  in their respective domains according to a reverse
lexicographic ordering (RLO) $\alpha \to i_\text{rlx}(\alpha)$, which
we recall to be defined by 
\begin{mydef}
Given $d\ge 1$ and  $\alpha,\beta\in\dN_0^d$, we say that $\alpha
\prec \beta$  if $|\alpha|<|\beta|$ or 
$|\alpha|=|\beta|$ and $\exists k\in\{1,\ldots,
d\}:\ \alpha_i=\beta_i$ for $i=1\ldots k-1$, and $\alpha_k<\beta_k$. 
\end{mydef}
\noindent An efficient indexing algorithm for this specific ordering
is discussed in \cite{deboor}, to which we refer the reader for
details. We see from the heat map that the coefficients
$\chi_{\alpha\gamma}$ decrease exponentially as $|\alpha|$
increases. This behaviour looks natural if one considers that
$V_\alpha$ is not positive definite on $\bar T_Q(s)$: its zeros are an
algebraic variety, whose structure becomes more and more complex as
$|\alpha|$ increases. The sign of $V_\alpha$ is important since
$\cD_\gamma(\bar\phi)\ge 0$ for $\bar\phi\in\bar T_Q(s)$, hence
$\chi_{\alpha\gamma}$ receives contributions of opposite signs from
adjacent domains separated by zeros of $V_\alpha$. For this reason, it
averages progressively to zero as $|\alpha|$ increases. The isolated
white points in the upper part of the plot correspond to values of
$\alpha$ and $\gamma$ for which $\chi_{\alpha\gamma}=0$; they have
been coloured as the lowest non--zero observed value of
$\log|\chi_{\alpha\gamma}|$, just to preserve the colour map.  

The reader could feel uncomfortable with the fact that the orthogonal
bases are not normalized on $\bar T_Q(s)$ such as $\cD_\gamma$
is. This is not really a problem as far as we are concerned, since
normalization constants change the rows of the RG
matrix by an irrelevant overall rescaling, as we shall see in the next
section. Moreover, normalizing the orthogonal polynomials
$\{V_\alpha\}$ and $\{U_\alpha\}$ as if they were probability densities
on $\bar T_Q(s)$ is not possible. Indeed, their integrals $Z_V$ and
$Z_U$, which are discussed in App.~A for the sake
of completeness, vanish for specific values of $\alpha$ and~$\kappa$  
due to non--positiveness.

\section{Ritz--Galerkin orthogonality}

The Fokker--Planck operator $\cLFP$ is a linear operator: given two
functions $\cP_1$, $\cP_2\in\cH$, with $\cH$ a sufficiently regular
function space on the $Q$--simplex, such that $\cLFP\cdot\cP_k=0$ for
$k=1,2$, then $\cLFP\cdot(a_1\cP_1+a_2\cP_2) = 0$ for any
$a_1,a_2\in\dR$. In other words, the solutions of the stationary FPE
belong to $\ker\cLFP\subset \cH$. Since in general $\dim\ker\cLFP>1$,
a specific solution can be singled out by imposing a set of additional
conditions. We shall come to this point in a while. For the time
being, we observe that   
\begin{equation}
\langle V,\cLFP\cdot \cP\rangle_\kappa = 0 \quad \forall\, V\in
\cH\quad \text{ if } \quad \cP\in\ker\cLFP\,. 
\label{eq:completesol}
\end{equation}
Conversely, a function $\cP\in\cH$ fulfilling $\langle V,\cLFP\cdot
\cP\rangle_\kappa = 0\ \ \forall\, V\in \cH$ is called \emph{a weak
  solution} of the FPE.  The idea underlying the RG approximation
method is to look for a weak solution by enforcing
eq.~(\ref{eq:completesol}) only for $V\in \bar\cH$, with $\bar\cH$ a
properly chosen subset of $\cH$. For instance, for $n\ge 1$, we could
opt for 
\begin{equation}
\bar\cH_n = {\rm span}\left\{\bar\phi^\alpha:\ \alpha\in\dN_0^{Q-1}
\text{  and } |\alpha|\le n\right\}\,. 
\end{equation}
Since $\{V_\alpha\}_{|\alpha|\le n}$ is a basis of $\bar\cH_n$, a RG
weak solution $\cP_n$ has to fulfill 
\begin{equation}
\langle V_\alpha,\cLFP\cdot \cP_n\rangle_\kappa = 0\,\qquad \forall\,
\alpha:\ |\alpha|\le n\,. 
\label{eq:projectFPE}
\end{equation}
As proved in a celebrated theorem by Lax and Milgram
\cite{laxmilgram}, a sufficient condition to make the search of weak
solutions (and therefore of RG weak solutions) a well--posed problem,
is that the following two properties are fulfilled: 
\begin{alignat}{4}
& \bullet \ \text{boundedness}& & \quad \rightsquigarrow & & \qquad
  \exists C<+\infty:\quad |\langle V,\cLFP\cdot V'\rangle| \le
  C\,||V|| ||V'||\,,\quad \forall\,
  V,V'\in\cH\,,\\[2.0ex] 
& \bullet \ \text{coerciveness}& & \quad \rightsquigarrow & & \qquad
  \exists c>0: \quad |\langle V,\cLFP\cdot V \rangle| \ge
  c||V||^2\,,\quad \forall\, V\in\cH\,, 
\end{alignat}
for some scalar product $\langle\,\cdot\,,\cdot\,\rangle$ on $\bar T_Q(s)$ (not necessarily $\langle\,\cdot\,,\,\cdot\,\rangle_\kappa$), with $||\cdot||^2 = \langle \cdot,\cdot\rangle$ being
the induced norm. Though it is
not difficult to check the boundedness condition for a $\cLFP$ with
polynomial coefficients $(A_k)_{k=1}^{Q-1}$ and $(B_{ik})_{i,k=1}^{Q-1}$ on a compact domain
such as $\bar T_Q(s)$, checking the coerciveness of $\cLFP$ is more
problematic, since this is related to the structure of the eigenvalue
spectrum of $(B_{ik})_{i,k=1}^{Q-1}$. We do not attempt any general proof in the
present paper. Instead, we adopt a heuristic approach where we just
apply the RG method to a given complex model and check out the
outcome. However, if the Lax--Milgram conditions are fulfilled, then
the C\'ea estimate 
\begin{equation}
||\cP - \cP_n|| \le \frac{C}{c}\inf_{\cQ_n\in\bar\cH_n}||\cP-\cQ_n||\,
\end{equation}
follows straightaway, stating that the RG solution $\cP_n$ is a {\it
  quasi}--best approximation on $\bar\cH_n$ to a truly weak solution
$\cP\in\cH$. In addition, the error $\cP-\cP_n$ is weakly orthogonal
to $\bar \cH_n$. That being said, we are ready to show how to adapt
the RG method to birth--death models with polynomial drift and
diffusion coefficients. 
\vskip 0.2cm
\noindent {$i$) If we expand $\cP_n$ according to
  eq.~(\ref{eq:polexp}) and insert the expansion into
  eq.~(\ref{eq:projectFPE}), we obtain 
\vskip -0.5cm
\begin{equation}
0 = \sum_{\gamma\in\Omega_n} \langle V_\alpha,\cLFP
\cdot\cD_\gamma\rangle_\kappa  \,c_\gamma  = \sum_{\gamma\in\Omega_n}
\psi_{\alpha\gamma}c_\gamma\,,\qquad \psi_{\alpha\gamma} \equiv
\langle V_\alpha,\cLFP 
\cdot\cD_\gamma\rangle_\kappa\,.
\label{eq:homogRG}
\end{equation}
Accordingly, the stationary FPE turns into a square homogeneous linear
system with coefficient matrix $\psi\in
\dR^{|\Omega_n|\times|\Omega_n|}$ and unknown vector
$c\in\dR^{|\Omega_n|}$, both indexed (for instance) via $\alpha \to
i_\text{rlx}(\alpha)$. If the problem is well posed, the eigenvalue
spectrum of $\psi$ must have a certain number of zeros $\nu_0>0$,
depending on $n$, $Q$ and the specific form of $(A_k)_{k=1}^{Q-1}$ and
$(B_{ik})_{i,k=1}^{Q-1}$. The linear system $\psi\cdot c=0$ must be augmented by
imposing that $c$ is normalized according to eq.~(\ref{eq:polnorm})
and by introducing a set of --~say~-- $Q\nbc$ additional equations to
enforce the boundary conditions, eq.~(\ref{eq:bc}). This leads us to a
larger non--homogeneous linear system $\Psi\cdot c = \eta$, of which
we know at present that 
\begin{alignat}{3}
\label{eq:systeq1}
& \Psi\in \dR^{(|\Omega_n|+1+Q\nbc)\times |\Omega_n|}\,: & &
\qquad\left\{\begin{array}{ll}\Psi_{ij} = \psi_{ij} & \hskip 0.12cm
\text{for } i,j=1,..,|\Omega_n|\,;\\[2.0ex] 
\Psi_{(|\Omega_n|+1)j} = 1 & \hskip 0.12cm \text{for }
j=1,\ldots,|\Omega_n|\,; 
\end{array}\right.\\[2.0ex] 
\label{eq:systeq2}
& \eta\in\dR^{(|\Omega_n|+1+Q\nbc)}\,: & & \qquad
\left\{\begin{array}{ll}\eta_{i} = 0 & \quad \text{for }
i=1,..,|\Omega_n|\,;\\[2.0ex] 
\eta_{|\Omega_n|+1} = 1\,. & \end{array}\right.
\end{alignat}
It should be observed that eq.~(\ref{eq:homogRG}) is left invariant by
any change of normalization of $V_\alpha$. Such a change would just
correspond to rescaling the rows of $\psi$.
\vskip 0.2cm
\noindent $ii$) Let us see how to set up the boundary conditions and
fill in the lowest $Q\nbc$ rows of $\Psi$ and elements
of~$\eta$. First, we recall that the stationary FPE can be written in
the form of a local conservation law, namely 
\begin{equation}
0 = \sum_{k=1}^{Q-1}\partial_k J_k(\bar\phi)\,, \qquad\qquad
J_k(\bar\phi) = -A_k(\bar\phi)\cP(\bar\phi) +
\frac{1}{2}\sum_{i=1}^{Q-1}\partial_i[B_{ik}(\bar\phi)\cP(\bar\phi)]\,, 
\label{eq:locFPE}
\end{equation}
where $J$ is naturally interpreted as a vector probability
current. Integrating both sides of eq.~(\ref{eq:locFPE}) over $\bar
T_Q(s)$ and making use of the divergence theorem yields 
\begin{equation}
0 = \int_{\partial \bar T_Q(s)}\rd \bar\phi \ \ \hat n(\bar\phi)\cdot
J(\bar\phi)\,,\qquad Q>2\,, 
\label{eq:zeroflux}
\end{equation}
with $\hat n(\bar\phi)$ representing the inward pointing unit vector
orthogonal to $\partial \bar T_Q(s)$ at $\bar\phi$. Clearly,
eq.~(\ref{eq:zeroflux}) means that there is no overall probability
flux across $\partial \bar T_Q(s)$ when the system is in
equilibrium. In order for eq.~(\ref{eq:bc}) to hold, the orthogonal
component of the probability current must vanish point-by-point on the
boundary (reflecting boundary conditions) and not just on average, \ie
the sought solution must fulfill $\hat n(\bar\phi)\cdot J(\bar\phi)=0$
for $\bar \phi\in\partial \bar T_Q(s)$. Unfortunately, this is a
continuous infinite set of conditions, which we however approximate by
a finite subset. To this end, we observe that $\partial \bar T_Q(s)$
is made of $Q$ $(Q-1)$--dimensional hypersurfaces, namely 
\begin{equation}
\partial\bar T_Q(s) = \bigcup_{k=1}^Q \fH_k\,, \qquad \left\{\begin{array}{ll}
 \fH_k\, = \{\bar\phi\in\bar T_Q(s):\quad \phi_k=0\}\,;\qquad &
 k=1,\ldots,Q-1\,,\\[2.0ex] 
 \fH_Q = \{\bar\phi\in\bar T_Q(s):\quad |\bar\phi|=s\}\,. & 
\end{array}\right.
\end{equation}
On each hypersurface $\fH_k$ we consider a regular grid of
zero--orthogonal--flux points $(\bar\phi_{km})_{m=1}^{\nbc}$ at
which we impose the condition $0= \hat n(\bar\phi_{km})\cdot
J(\bar\phi_{km})$, \ie 
\begin{alignat}{3}
\label{eq:bcorth}
& 0 =  J_k(\bar\phi_{km}) \,,& & \qquad k = 1,\ldots,Q-1\, \quad
m=1,\ldots,\nbc\,,\\[1.0ex] 
& 0 = \sum_{k=1}^{Q-1}J_k(\bar\phi_{Qm})\,, & & \qquad
m=1,\ldots,\nbc\,. 
\label{eq:bcskew}
\end{alignat}
An illustrative example corresponding to $Q=3$ and $\nbc=10$ is shown
in Fig.~\ref{fig:bc}, where the boundary points have been chosen
according to 
\begin{equation}
\bar\phi_{1m} = \left(0,s\frac{m}{\nbc+1}\right)\,,\quad 
\bar\phi_{2m} = \left(s\frac{m}{\nbc+1},0\right)\,,\quad
\bar\phi_{3m} =\left(s\frac{m}{\nbc+1},s\frac{\nbc+1-m}{\nbc+1} \right)\,.
\end{equation}
\begin{figure}[t!]
    \centering
    \includegraphics[width=0.6\textwidth]{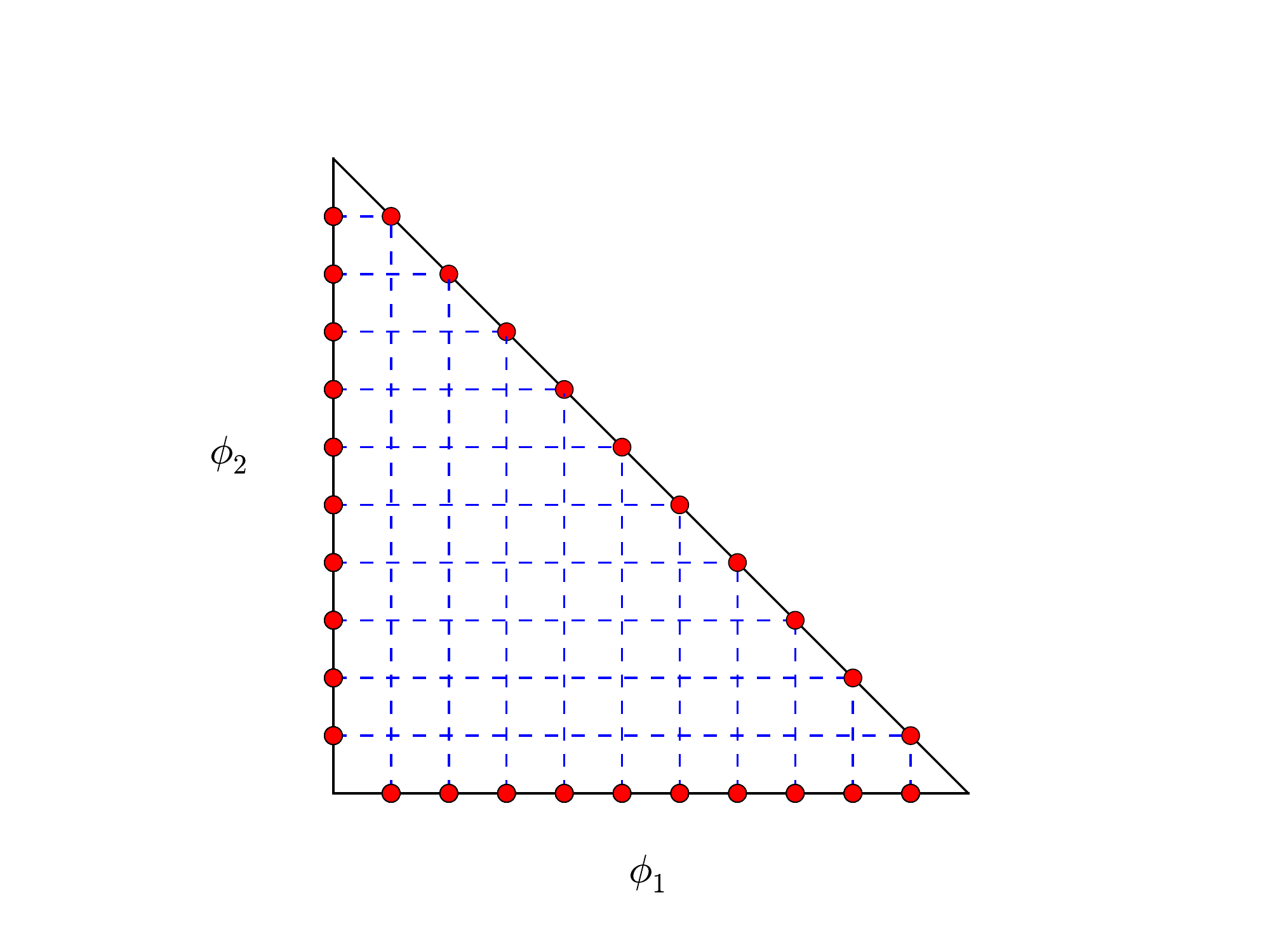}
    \vskip -0.2cm
    \caption{ \footnotesize Zero--orthogonal--flux points for $Q=3$ and $\nbc=10$.}
  \label{fig:bc}
\end{figure}

\noindent $iii$) As already observed, the rank $r_\psi$ of $\psi$ is
expected not to be maximal, \ie
$r_\psi=|\Omega_n|-\nu_0<|\Omega_n|\equiv r_{\psi,\text{max}}$. The
normalization condition eq.~(\ref{eq:polnorm}) adds a linearly
independent row to the system $\psi\cdot c=0$, thus increasing the
rank of the coefficient matrix by one. Each additional boundary
condition adds another linearly independent row and 
further increases the rank of the coefficient matrix until this
becomes maximal. From this point on, \ie for $Q\nbc>\nu_0-1$, the rank
of the coefficient matrix keeps maximal, while the system becomes
overconstrained and thus inconsistent (to understand this, imagine to
perform a row echelon reduction of the system $\Psi\cdot c = \eta \to
\Psi_\text{red}\cdot c=\eta_\text{red}$; the reduced row echelon form
$\Psi_\text{red}$ has still maximal rank $r_\Psi=|\Omega_n|$; its last
$Q\nbc-\nu_0+1$ rows are full of zeros, while in general the last
$Q\nbc-\nu_0+1$ elements of $\eta_\text{red}$ are expected not to
vanish). This is particularly inconvenient, as it compels us to  very
carefully choose an exact number $Q\nbc = \nu_0-1$ of boundary
points. Though reasonable, we have no theoretical argument to prove
that $\nu_0-1 \propto Q$. We follow a different approach: an
alternative is indeed to impose an arbitrary number $Q\nbc>\nu_0-1$ of
boundary conditions and consider the normal system  
\begin{equation}
(\trans{\Psi}\cdot\Psi)\cdot c = \trans{\Psi}\cdot \eta\,.
\label{eq:normsyst}
\end{equation}
in place of the original one (least--squares
problem). Eq.~(\ref{eq:normsyst}) is consistent for any choice of
$Q\nbc>\nu_0-1$. Indeed, since $\Psi$ has maximal rank,
$\trans{\Psi}\cdot\Psi\in \dR^{|\Omega_n|\times |\Omega_n|}$ has no
zero eigenvalues, hence it can be inverted. The system is consistent
as $(\trans{\Psi}\cdot\eta)_k = 1$  for all
$k=1,\ldots,|\Omega_n|+1+Q\nbc$, as a consequence of
eqs.~(\ref{eq:systeq1})--(\ref{eq:systeq2}). Clearly, the original
system and eq.~(\ref{eq:normsyst}) are not equivalent.  

Notice that $\Psi$ is in general expected to have a large condition
number (the ratio between its largest and lowest singular value) and
the latter is expected to get larger as $n$ increases. Since the
condition number of $\trans{\Psi}\cdot\Psi$ is the square of the
condition number of $\Psi$, the inversion of $\trans{\Psi}\cdot\Psi$
might be computationally critical. Therefore, an appropriate inversion
algorithm should be used in order to solve eq.~(\ref{eq:normsyst}). We
use the CGNR algorithm in our numerical tests, see
ref.~\cite[chapt.~8]{saad} for details.    

 The effects of imposing more and more boundary conditions will be
 discussed in a specific example in sect.~5. We can  say in advance
 that the condition number of $\trans{\Psi}\cdot \Psi$ is not
 sensitive to $\nbc$ and that $\cP_n$ rapidly converges as $\nbc$ increases. 
\vskip 0.2cm 
\noindent $iv$) We need to discuss how to concretely work out and
compute the matrix elements $\psi_{\alpha\gamma}$. Here, the
assumption that $(A_k)_{k=1}^{Q-1}$ and $(B_{ik})_{i,k=1}^{Q-1}$ are polynomials becomes
practically decisive. Indeed, we observe that $\partial_k^m\cD_\gamma$
is a polynomial on $\bar T_Q(s)$ with $\deg\{\partial_k^m\cD_\gamma\}
= |\slashed{\gamma}|-m$ for $m\le |\slashed{\gamma}|$, while
$\phi_k^m\cD_\gamma$ is a polynomial on $\bar T_Q(s)$ with
$\deg\{\phi_k^m\cD_\gamma\} = |\slashed{\gamma}|+m$. Since
$\{\cD_\gamma\}$ is a polynomial basis, it must be possible to express
both $\partial_k^m\cD_\gamma$ and $\phi_k^m\cD_\gamma$ as linear
combinations of some $\{\cD_{\gamma'}\}$. Now, since $\cLFP$ is a second
order partial differential operator, we never need to differentiate
more than twice. Analogously, since $\cLFP$ is usually derived from a
Master Equation resulting from a detailed balance, $(A_k)_{k=1}^{Q-1}$ and
$(B_{ik})_{i,k=1}^{Q-1}$ are usually not more than quadratic polynomials (this
statement is of course less universal -- as the reader may
understand~-- since transition rates depend on the specific model, but
is often true). Instead of writing a general formula to expand
$\phi_i^p\phi_k^q\partial^r_m\partial^s_n\cD_\gamma(\bar\phi)$ as a
linear combination of Dirichlet distributions, we prefer to report
formulae for specific choices of indices and exponents. To this
aim, we need to introduce some additional notation. We define 
\begin{equation}
\gamma_{\ell^\pm} \equiv (\gamma_1,\ldots,\gamma_{\ell-1},\gamma_\ell\pm 1,\gamma_{\ell+1},\ldots,\gamma_Q)\,.
\end{equation}
Similarly, we define $\gamma_{\ell^+m^+}$, $\gamma_{\ell^+m^-}$,
$\gamma_{\ell^{++}}$, \etc. as results of the iterated application of
index--raising operators $\oplus_\ell\cdot \gamma \equiv \gamma_{\ell^+}$
and index--lowering operators $\ominus_\ell \cdot \gamma \equiv
\gamma_{\ell^-}$, somewhat similar to the creation and destruction
operators of the quantum harmonic oscillator. Based on this, reference
formulae read
\vskip 0.3cm
\hrule
\vskip 0.2cm
\begin{equation}
\phi_\ell\cD_\gamma(\bar\phi) = s\frac{\gamma_\ell}{|\gamma|}\cD_{\gamma_{\ell^+}}(\bar\phi)\,,
\label{eq:firstalg}
\end{equation}
\begin{equation}
\partial_\ell\cD_\gamma(\bar\phi) = s^{-1}(|\gamma|-1)[\theta_{\gamma_\ell,2}\cD_{\gamma_{\ell^-}}(\bar\phi) - \theta_{\gamma_Q,2}\cD_{\gamma_{Q^-}}(\bar\phi)]\,,
\label{eq:delDgamma}
\end{equation}
\begin{equation}
\phi_\ell\partial_\ell \cD_\gamma(\bar\phi) = \theta_{\gamma_\ell,2}(\gamma_\ell-1)\cD_\gamma(\bar\phi) - \theta_{\gamma_Q,2}\gamma_\ell \cD_{\gamma_{\ell^+Q^-}}(\bar\phi)\,,
\label{eq:phidelDgamma}
\end{equation}
\begin{equation}
\phi^2_\ell\partial_\ell \cD_\gamma(\bar\phi) =\frac{s}{|\gamma|}\left\{\theta_{\gamma_\ell,2}\gamma_\ell(\gamma_\ell-1)\cD_{\gamma_{\ell^+}}(\bar\phi) - \theta_{\gamma_Q,2}\gamma_\ell(\gamma_\ell+1) \cD_{\gamma_{\ell^{++}Q^-}}(\bar\phi)\right\}\,,
\end{equation}
\begin{equation}
\phi_\ell\partial_m \cD_\gamma(\bar\phi)|_{\ell\ne m} = \gamma_\ell\left[\theta_{\gamma_m,2}\cD_{\gamma_{\ell^+m^-}}(\bar\phi) - \theta_{\gamma_Q,2} \cD_{\gamma_{\ell^+Q^-}}(\bar\phi)\right]\,,
\end{equation}
\vskip -0.2cm
\begin{align}
\partial_\ell^2\cD_\gamma(\bar\phi) & = s^{-2}(|\gamma|-1)(|\gamma|-2)\cdot \left\{\theta_{\gamma_\ell,3}\cD_{\gamma_{\ell^{--}}}(\bar\phi) \right.\nonumber\\[1.0ex]
& \left. - 2\theta_{\gamma_\ell,2}\theta_{\gamma_Q,2}\cD_{\gamma_{\ell^-Q^-}}(\bar\phi) + \theta_{\gamma_Q,3}\cD_{\gamma_{Q^{--}}}(\bar\phi)\right\}\,,
\end{align}
\vskip -0.2cm
\begin{align}
\phi_\ell\partial_\ell^2\cD_\gamma(\phi) & = s^{-1}(|\gamma|-1)\cdot \left\{(\gamma_\ell-2)\theta_{\gamma_\ell,3}\cD_{\gamma_{\ell^-}}(\bar\phi)\right.\nonumber\\[1.0ex]
& \hskip -0.5cm  \left.-2(\gamma_\ell-1)\theta_{\gamma_\ell,2}\theta_{\gamma_Q,2}\cD_{\gamma_{Q^-}}(\bar\phi) +\gamma_\ell\theta_{\gamma_Q,3}\cD_{\gamma_{\ell^+Q^{--}}}(\bar\phi)\right\}\,,
\end{align}
\vskip -0.2cm
\begin{align}
\phi_m\partial_\ell^2\cD_\gamma(\bar\phi)|_{\ell\ne m} & = s^{-1}\gamma_m(|\gamma|-1)\cdot \left\{\theta_{\gamma_\ell,3}\cD_{\gamma_{m^+\ell^{--}}}(\bar\phi)\right.\nonumber\\[1.0ex]
& \hskip -1.38cm  \left.-2\theta_{\gamma_\ell,2}\theta_{\gamma_Q,2}\cD_{\gamma_{m^+\ell^-Q^-}}(\bar\phi) +\theta_{\gamma_Q,3}\cD_{\gamma_{m^+Q^{--}}}(\bar\phi)\right\}\,,
\end{align}
\vskip -0.2cm
\begin{align}
\phi_\ell^2\partial_\ell^2\cD_\gamma(\bar\phi) & = (\gamma_\ell-2)(\gamma_\ell-1)\theta_{\gamma_\ell,3}\cD_{\gamma}(\bar\phi) - 2(\gamma_\ell-1)\gamma_\ell\theta_{\gamma_\ell,2}\theta_{\gamma_Q,2}\cD_{\gamma_{\ell^+Q^-}}(\bar\phi) \nonumber\\[1.0ex]
& \hskip -0.4cm +\gamma_\ell(\gamma_\ell+1)\theta_{\gamma_Q,3}\cD_{\gamma_{\ell^{++}Q^{--}}}(\bar\phi)\,,
\end{align}
\vskip -0.2cm
\begin{align}
\phi_m^2\partial_\ell^2\cD_\gamma(\bar\phi)|_{\ell\ne m} & = \gamma_m(\gamma_m+1)\left\{\theta_{\gamma_\ell,3}\cD_{\gamma_{m^{++}\ell^{--}}}(\bar\phi)\right. \nonumber\\[1.0ex]
& \hskip -1.26cm \left. - 2\theta_{\gamma_\ell,2}\theta_{\gamma_Q,2}\cD_{\gamma_{m^{++}\ell^-Q^-}}(\bar\phi)
+\theta_{\gamma_Q,3}\cD_{\gamma_{m^{++}Q^{--}}}(\bar\phi)\right\}\,,
\end{align}
\vskip -0.2cm
\begin{align}
\phi_\ell\phi_m\partial_\ell^2\cD_\gamma(\bar\phi)|_{\ell\ne m} & =
\gamma_m\left\{(\gamma_\ell-2)\theta_{\gamma_\ell,3}\cD_{\gamma_{m^{+}\ell^{-}}}(\bar\phi)\right. \nonumber\\[1.0ex] 
& \hskip -1.76cm \left. -
2(\gamma_\ell-1)\theta_{\gamma_\ell,2}\theta_{\gamma_Q,2}\cD_{\gamma_{m^{+}Q^-}}(\bar\phi) 
+\gamma_\ell\theta_{\gamma_Q,3}\cD_{\gamma_{m^{+}\ell^+Q^{--}}}(\bar\phi)\right\}\,, 
\end{align}
\vskip -0.2cm
\begin{align}
\partial_\ell\partial_m\cD_\gamma(\bar\phi)|_{\ell\ne m} & = s^{-2}(|\gamma|-1)(|\gamma|-2)\nonumber\\[1.0ex]
& \hskip -1.15cm \cdot
\left[\theta_{\gamma_\ell,2}\theta_{\gamma_m,2}\cD_{\gamma_{(\ell^-)(m^-)}}(\bar\phi)
  -\theta_{\gamma_\ell,2}\theta_{\gamma_Q,2}\cD_{\gamma_{(\ell^-)(Q^-)}}(\bar\phi)\right.\nonumber\\[1.0ex] 
& \left. \hskip -1.11cm -\theta_{\gamma_m,2}\theta_{\gamma_Q,2}
  \cD_{\gamma_{(m^-)(Q^-)}}(\bar\phi) 
 + \theta_{\gamma_Q,3}\cD_{\gamma_{Q^{--}}}(\bar\phi)\right]\,, 
\end{align}
\vskip -0.2cm
\begin{align}
\phi_\ell\partial_\ell\partial_m\cD_\gamma(\bar\phi)|_{\ell\ne m} & = s^{-1}(|\gamma|-1)\nonumber\\[1.0ex]
& \hskip -1.55cm \cdot \left\{(\gamma_\ell-1)
\left[\theta_{\gamma_\ell,2}\theta_{\gamma_m,2}\cD_{\gamma_{m^-}}(\bar\phi)-
  \theta_{\gamma_\ell,2}\theta_{\gamma_Q,2}\cD_{\gamma_{Q^-}}(\bar\phi)\right]\right. \nonumber\\[1.0ex] 
&
\hskip -1.55cm
\left.-\gamma_\ell\left[\theta_{\gamma_m,2}\theta_{\gamma_Q,2}\cD_{\gamma_{\ell^+m^-Q^-}}(\bar\phi)
  -
  \theta_{\gamma_Q,3}\cD_{\gamma_{\ell^+Q^{--}}}(\bar\phi)\right]\right\}\,, 
\end{align}
\vskip -0.2cm
\begin{align}
\phi_\ell\phi_m\partial_\ell\partial_m\cD_\gamma(\bar\phi)|_{\ell\ne m} & =\nonumber\\[1.0ex] 
& \hskip -2.0cm
(\gamma_\ell-1)(\gamma_m-1)\theta_{\gamma_\ell,2}\theta_{\gamma_m,2}\cD_\gamma(\bar\phi)
-
\gamma_m(\gamma_\ell-1)\theta_{\gamma_\ell,2}\theta_{\gamma_Q,2}\cD_{\gamma_{m^+Q^-}}(\bar\phi)\nonumber\\[1.0ex] 
& \hskip -2.0cm -
\gamma_\ell(\gamma_m-1)\theta_{\gamma_m,2}\theta_{\gamma_Q,2}\cD_{\gamma_{\ell^+Q^-}}(\bar\phi)
+
\gamma_\ell\gamma_m\theta_{\gamma_Q,3}\cD_{\gamma_{\ell^+m^+Q^{--}}}(\bar\phi)\,, 
\label{eq:lastalg}
\end{align}
\vskip 0.2cm
\hrule
\vskip 0.3cm
\noindent where 
\begin{equation}
\theta_{a,b} = \left\{\begin{array}{ll}1 & \quad \text{if}\quad a \ge b\,,\\[2.0ex]
0 & \quad \text{otherwise}\,.\end{array}\right.
\end{equation}
Now, projecting -- by way of example -- eq.~(\ref{eq:delDgamma}) onto $V_\alpha$ yields
\begin{equation}
\langle V_\alpha, \partial_\ell \cD_\gamma\rangle_\kappa =
s^{-1}(|\gamma|-1)[\theta_{\gamma_\ell,2}\,\chi_{\alpha\gamma_{\ell^-}}
  - \theta_{\gamma_Q,2}\,\chi_{\alpha\gamma_{Q^-}}]\,. 
\end{equation}
Analogously it be can done for all
eqs.~(\ref{eq:firstalg})--({\ref{eq:lastalg}); we see indeed that
  projecting the whole function $\cLFP\cdot\cP_n$ onto $V_\alpha$ is
  just a matter of tedious yet simple algebra. We conclude that
  $\psi_{\alpha\gamma}$ can be expanded as a self--contained sum of
  contributions, each being proportional to some matrix element
  of~$\chi$. However, we observe that 
\begin{equation}
  \left[\oplus_1^{m_1}\ldots\oplus_{Q-1}^{m_{Q-1}}\right]\cdot
  \left[\ominus_1^{\ell_1}\ldots\ominus_{Q-1}^{\ell_{Q-1}}\right]\cdot\gamma\,\in\Omega_{n+\sum_km_k-\sum_k\ell_k}\quad
  \text{ for } \gamma\in\Omega_n\, \text{ and }\ \{\ell_k<\gamma_k\}\,.  
\end{equation}
If $\max_k\deg\{A_k\} = m_A$, then the action of the drift term on
$\cP_n$ mixes Dirichlet distributions with index arrays in the \emph{bucket spaces}
$\Omega_{n-1},\Omega_n,\ldots,\Omega_{n+m_A-1}$. Likewise, if
$\max_{i,k}\deg\{B_{ik}\}=m_B$, then the action of the diffusion term
on $\cP_n$ mixes Dirichlet distributions with index arrays in the \emph{bucket spaces}
$\Omega_{n-2},\Omega_{n-1},\ldots,\Omega_{n+m_B-2}$. Accordingly,
in order to compute the matrix $\psi$, we need to compute
$\chi_{\alpha\gamma}$ for $|\alpha|\le n$ and for $\gamma\in\Omega_k$ for some 
$k\in\{n-2,\ldots,n_\text{max}\}$, with $n_\text{max} = \max\{n+m_A-1,n+m_B-2\}$. 
\vskip 0.2cm
\noindent $v$) In order to work out
eqs.~(\ref{eq:bcorth})--(\ref{eq:bcskew}), we first insert
eq.~(\ref{eq:polexp}) into $J_k$  and extract the coefficient
multiplying each $c_\gamma$, namely 
\begin{equation}
J_k(\bar\phi) =
\sum_{\gamma\in\Omega_n}c_\gamma\left\{A_k(\bar\phi)\cD_\gamma(\bar\phi)
-
\frac{1}{2}\sum_{i=1}^{Q-1}\partial_i[B_{ik}(\bar\phi)\cD_\gamma(\bar\phi)]\right\}
\equiv
\sum_{\gamma\in\Omega_n}\Upsilon_{k\gamma}(\bar\phi)c_\gamma\,. 
\end{equation}
We need to compute each matrix coefficient
$\Upsilon_{k\gamma}(\bar\phi)$ just for two sets of boundary points, namely
$\Upsilon_{k\gamma}(\bar\phi_{k,m})$ (in order to impose the boundary
conditions on $\fH_k$) and $\Upsilon_{k\gamma}(\bar\phi_{Q,m})$ (in
order to impose the boundary conditions on $\fH_Q$). The reader should
notice that $\cD_\gamma(\bar\phi_{km})=0$ unless $\gamma_k=1$ as well
as $\cD_\gamma(\bar\phi_{Qm})=0$ unless $\gamma_Q=1$. Since
$\Upsilon_{k\gamma}$ depends on both $\cD_\gamma$ and
$\{\partial_i\cD_\gamma\}$, this means that
$\Upsilon_{k\gamma}(\bar\phi_{km})=0$ unless $\gamma_k=1,2$ and
equally $\Upsilon_{k\gamma}(\bar\phi_{Qm})=0$ unless
$\gamma_Q=1,2$. Therefore, we conclude that the only unknowns taking
part in the boundary equations are those $c_\gamma$ which have at
least one component $\gamma_k=1,2$ with $k=1,\ldots,Q$.  
\vskip 0.6cm 
\begin{myremark}
By now, it should be sufficiently clear what the pros and cons of
projecting $\cP$ onto a set of Dirichlet distributions are. We find
it worthwhile summarizing them: 
\begin{itemize}
\item{the Dirichlet distributions $\{\cD_\gamma\}_{\gamma\in\Omega_n}$
  are not orthogonal polynomials with respect to the scalar product
  $\langle\cdot,\cdot\rangle_\kappa$, yet they form a basis of $\bar
  \cH_n$;} 
\item{while the zeros of the orthogonal polynomials $\{V_\alpha\}$ and
  $\{U_\alpha\}$ are non--trivial algebraic varieties, the Dirichlet
  distributions  $\{\cD_\gamma\}_{\gamma\in\Omega_n}$ are
  non--negative on $\bar T_Q(s)$. This means that the positiveness of
  $\cP_n$ relies entirely on the signs of the expansion coefficients
  $\{c_\gamma\}$. If $c_\gamma\ge 0$ $\ \forall\gamma\in\Omega_n$,
  then $\cP_n$ can be statistically interpreted as a distributional
  mixture;} 
\item{if for too small values of $n$ the RG approximation gives
  $\cP_n(\bar\phi)<0$ for $\bar\phi$ in some positive--measure subset
  of $\bar T_Q(s)$, it is anyway possible to obtain a decent
  (non--quasi--best) approximation of $\cP$ by changing the sign of
  some coefficient $c_\gamma$ and by subsequently renormalizing the
  whole vector $c$;} 
\item{the differentiation rules of $\cD_\gamma$ generate
  self--contained algebraic expressions involving Dirichlet
  distributions with different indices. Although $\psi$ is a dense
  matrix, it can be easily computed. Notice, however, that not only
  $|\Omega_n|$ inflates almost exponentially with $Q$, but also the
  CPU time needed to compute $\chi_{\alpha\gamma}$ for a given pair
  $(\alpha,\gamma)$ blows up, since eq.~(\ref{eq:scalprod}) contains a
  non--factorizable multiple sum $\sum_{\beta\le\alpha} =
  \sum_{\beta_1=0}^{\alpha_1}\ldots
  \sum_{\beta_{Q-1}=0}^{\alpha_{Q-1}}$\,;} 
\item{the Dirichlet distribution $\cD_\gamma$ vanishes on $\fH_k$
  unless $\gamma_k=1$. It is therefore very simple to keep track of
  which terms are responsible for the behaviour of $\cP_n$ on
  $\partial \bar T_Q(s)$. Such a task would be a nightmare with any
  other polynomial basis. \qed} 
\end{itemize}
\vskip 0.2cm
\end{myremark}

\section{Example~1: binary voter model with zealots}

The binary voter model, introduced in \cite{Clifford1973,Holley1975},
can be considered as an archetype of agent--based models for opinion
dynamics. Owing  popularity to its exact solvability on a lattice in
any dimension, the model has been studied in a number of variants. We
refer the reader to~\cite{FortCastLor} for a comprehensive review of
the relevant literature. The microscopic dynamics of the model is
simply defined. Agents carry a binary variable $v\in\{+1,-1\}$ and are
selected at random for transitions. When an agent is selected, she
flips her variable to that of a neighbour agent, also chosen at
random. In a certain time the system collapses to a consensus state
(all agents eventually share the same opinion), unless a stabilization
mechanism is turned on. One possibility is to perturb the system by
introducing {\it zealots} among the agents, \ie special individuals
who never change their opinion. Zealots in the context of the binary
voter model have been originally proposed in \cite{Mobilia2003}. If
competing zealots with opposite opinions are present, consensus states
are prevented as discussed in \cite{Mobilia2007}. As far as we are
concerned here, the binary voter model with  zealots is of interest
because 
\begin{itemize}
\item{it is a one--dimensional model, \ie $Q=2$;}\\[-2.0ex]
\item{the FPE of the model can be solved exactly.}
\end{itemize}
Both these features make it a simple case study to test the RG
method. Let $N$ denote the total number of agents, $Z_\pm$ the number
of zealots with $v=\pm 1$ and $N_\pm$ the number of dynamic agents
with $v=\pm 1$. Along with \cite{Mobilia2007}, we define $\phi=N_+/N$,
$z_\pm = Z_\pm/N$ and $s=1-z_+-z_-$. Accordingly, it must be
$0\le\phi\le s$, \ie $\bar T_Q(s)$ is just an interval in this
case. The FPE of the model reads 
\begin{equation}
0 = -{\partial_\phi}[A(\phi)P(\phi)] +
\frac{1}{2}{\partial^2_\phi}[B(\phi)P(\phi)] =
\partial_\phi\left\{-A(\phi)P(\phi) +
\frac{1}{2}{\partial_\phi}[B(\phi)P(\phi)]\right\} = \partial_\phi
J(\phi)\,, 
\label{eq:bvmfpe}
\end{equation}
with $\partial_\phi=\partial/\partial\phi$. The drift and diffusion
coefficients are given by 
\begin{align}
A(\phi) & = [z_+s - \phi(1-s)]\,,\\[1.0ex]
B(\phi) & = N^{-1}[(\phi+z_+)(s-\phi) + \phi(s+z_--\phi)]\,.
\end{align}
If one introduces the auxiliary variables $\delta=z_+-z_-$,
$r=\sqrt{\delta^2+4s}$ and $u_\pm = s/2-\delta/4\pm r/4$, then the
exact solution of the FPE \cite{Mobilia2007} reads 
\begin{equation}
\cP(\phi) =
W\cdot[(\phi-u_+)(\phi-u_-)]^{(Z_++Z_--2)/2}\left[1+\frac{r}{2\phi-s-\frac{r-\delta}{2}}\right]^{(\delta/2r)(2N-Z_+-Z-)}\,, 
\label{vmsol}
\end{equation}
with $W$ being a normalization constant such that
$\int_0^s\rd\phi\,\cP(\phi)=1$. When $z_\pm=z$, the solution collapses
to 
\begin{equation}
\cP_\text{sym}(\phi) = W\cdot[zs+2\phi(s-\phi)]^{Nz-1}\,.
\end{equation}
Notice that $\cP_\text{sym}$ is a polynomial with
$\deg\{\cP_\text{sym}\}=2Nz-2$, while $\cP$ is a rational function for
$z_+\ne z_-$. It should be observed that the condition $\hat
n(\bar\phi)\cdot J(\bar\phi)=0$ is meaningless for $Q=2$ since $\hat
n$ is not defined at all. Indeed, $Q=2$ is a degenerate case:
$J(\phi)=\text{const.}$ is a first integral of eq.~(\ref{eq:bvmfpe})
and eq.~(\ref{eq:bc}) is simply fulfilled provided we choose the
constant to be zero. Now, a Dirichlet distribution with $Q=2$ is
actually a beta distribution 
\begin{equation}
\cD_{(\gamma_1,\gamma_2)}(\phi) =
\frac{\Gamma(\gamma_1+\gamma_2)}{\Gamma(\gamma_1)\Gamma(\gamma_2)}s^{1-\gamma_1-\gamma_2}\phi^{\gamma_1-1}(s-\phi)^{\gamma_2-1}
= B_{\gamma_1\gamma_2}(\phi)\,, 
\end{equation}
and the \emph{bucket space} amounts in this case to
\begin{equation}
\Omega_n =
\biggl\{(n+1,1),\,(n,2),\ldots,(2,n),\,(1,n+1)\biggr\}\,,
\qquad |\Omega_n| = {n+1\choose n} = n+1\,. 
\end{equation}
\begin{figure}[t!]
  \centering
  \includegraphics[width=0.95\textwidth]{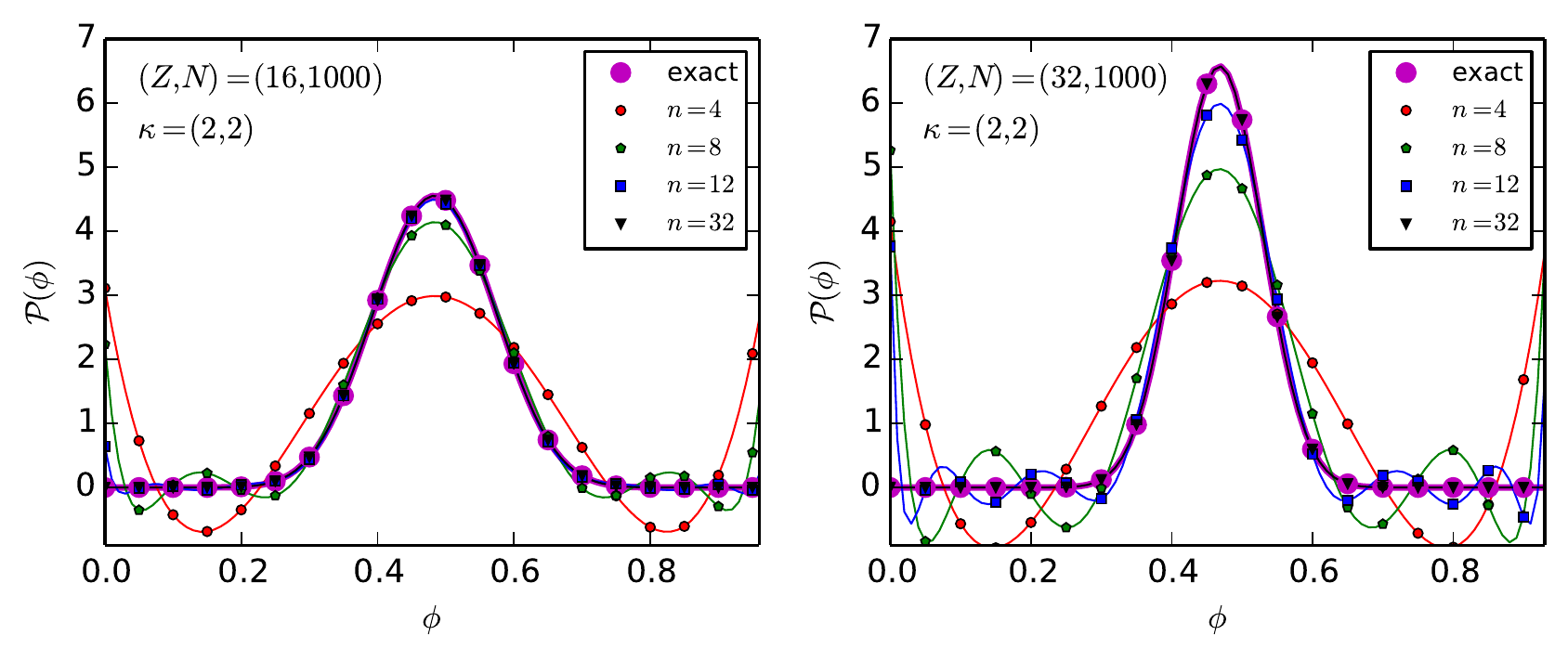}
  \vskip 0.9cm
  \includegraphics[width=0.95\textwidth]{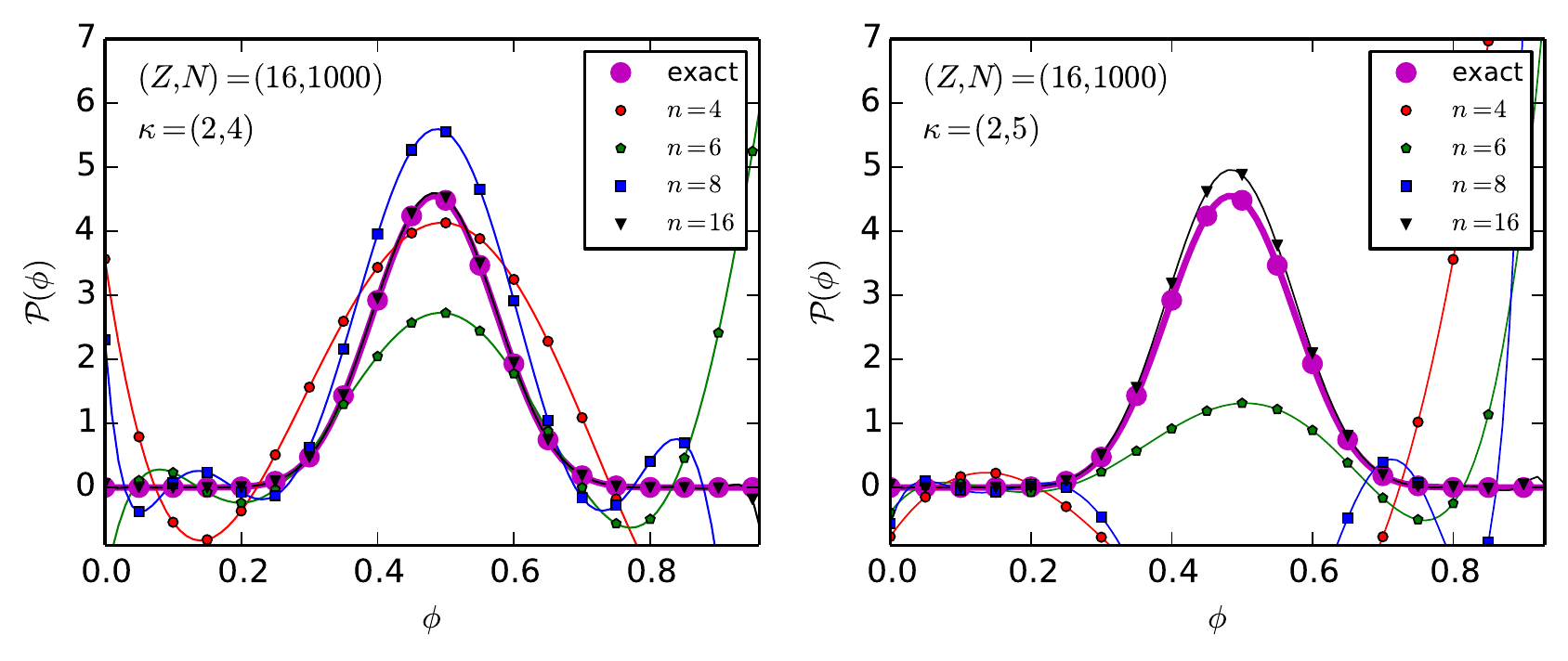}
  \vskip 0.9cm
  \includegraphics[width=0.95\textwidth]{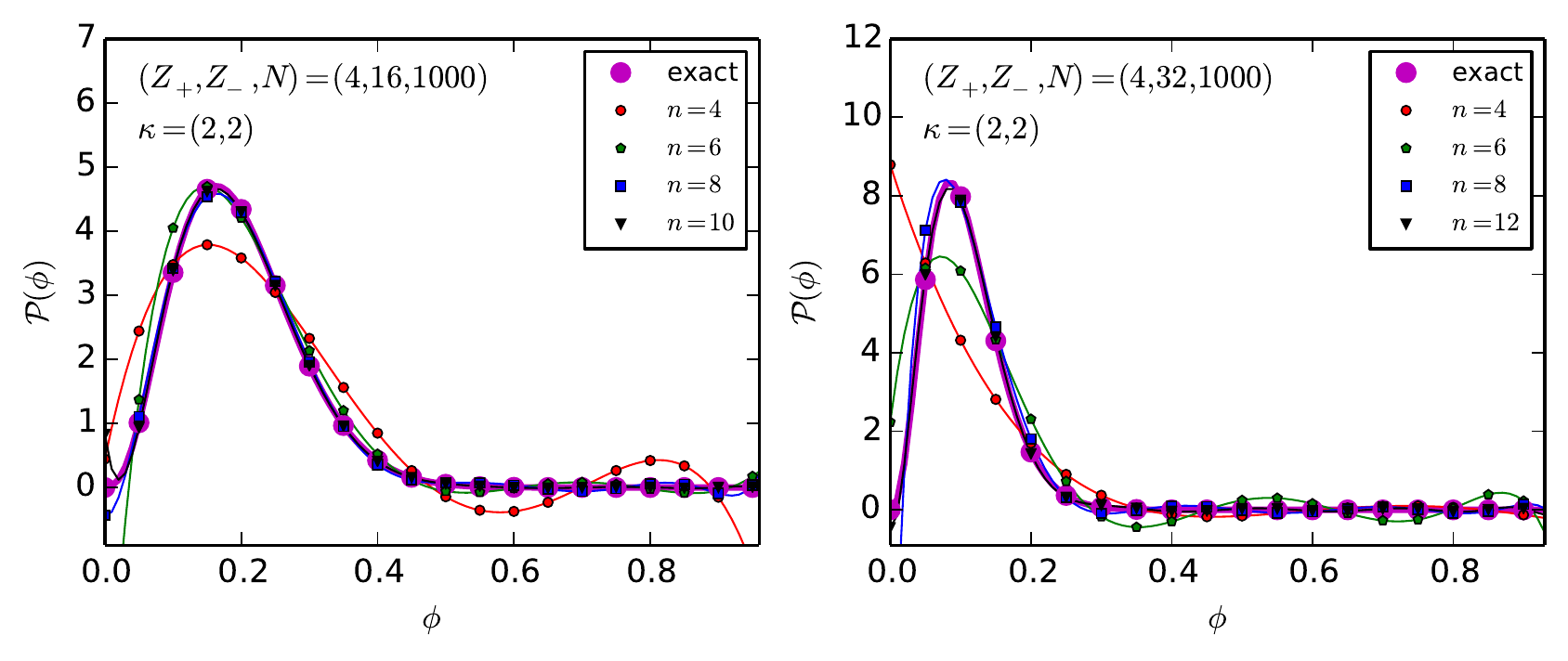}
\vskip -0.0cm
\caption{ \footnotesize RG approximations of the probability density
  of the binary voter model for various parameter sets.} 
\label{fig:bvmone}
\end{figure}

\begin{figure}[t!]
  \centering
  \hskip -0.5cm    
  \includegraphics[width=0.95\textwidth]{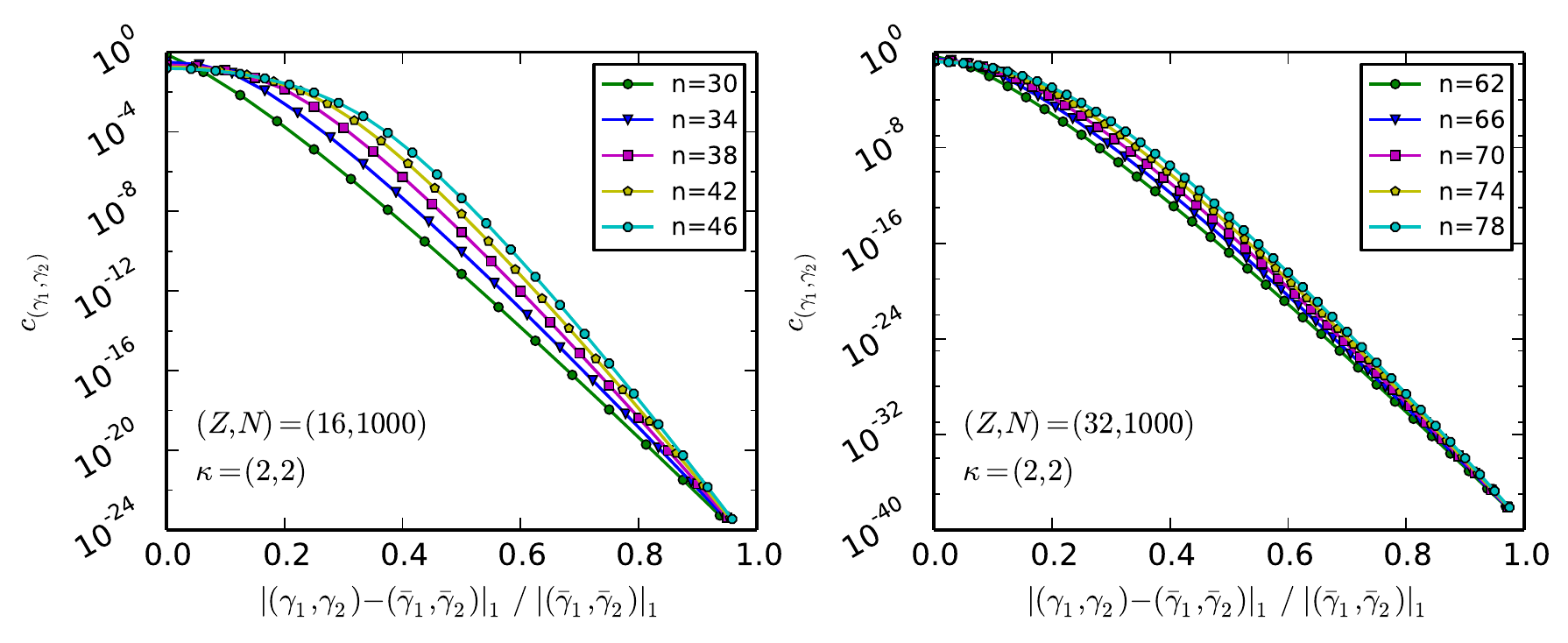}
\vskip 0.1cm
\caption{ \footnotesize Dirichlet spectra of the RG approximations.}
\label{fig:bvmspectra}
\end{figure}
\vskip -0.4cm
By using the differentiation formulae reported in sect.~3 and some
scratch paper, we can easily work out the matrix coefficients
$\psi_{\alpha\gamma} = \langle V_\alpha, -A\cD_\gamma +
2^{-1}\partial_\phi[B\cD_\gamma]\rangle_\kappa$, namely 
\begin{align}
\langle V_\alpha, -A\cD_\gamma\rangle_\kappa & =
-z_+s\,\chi_{\alpha(\gamma_1,\gamma_2)}
+s(1-s)\frac{\gamma_1}{\gamma_1+\gamma_2}\,\chi_{\alpha(\gamma_1+1,\gamma_2)}\,, \end{align} 
and
\begin{align}
\langle V_\alpha,\frac{1}{2}\partial_\phi[B\cD_\gamma]\rangle_\kappa & = \frac{1}{N}\biggl\{
-2s\frac{\gamma_1}{\gamma_1+\gamma_2}\,\chi_{\alpha(\gamma_1+1,\gamma_2)}
+\frac{1}{2}(2s-z_++z_-)\,\chi_{\alpha(\gamma_1,\gamma_2)}\nonumber\\[3.0ex]
& \hskip -1.6cm
-\frac{s}{\gamma_1+\gamma_2}\left[\gamma_1(\gamma_1-1)\theta_{\gamma_1,2}\,\chi_{\alpha(\gamma_1,\gamma_2)}
  -\gamma_1(\gamma_1+1)\theta_{\gamma_2,2}\,\chi_{\alpha(\gamma_1+1,\gamma_2-1)}\right]\nonumber
\end{align}
\begin{align}
& \hskip 1.0cm
+\frac{1}{2}(2s-z_++z_-)\left[(\gamma_1-1)\theta_{\gamma_1,2}\,\chi_{\alpha(\gamma_1,\gamma_2)}
  -
  \gamma_1\theta_{\gamma_2,2}\,\chi_{\alpha(\gamma_1+1,\gamma_2-1)}\right]\nonumber\\[3.0ex] 
& \hskip 1.0cm
+\frac{z_+}{2}(\gamma_1+\gamma_2-1)\left[\theta_{\gamma_1,2}\chi_{\alpha(\gamma_1-1,\gamma_2)}
-\theta_{\gamma_2,2}\chi_{\alpha(\gamma_1,\gamma_2-1)}\right]\biggr\}\,,
\end{align}
with $\gamma\in\Omega_n$ and $\alpha=0,1,\ldots,n$. Having coded
$\psi$, we checked numerically that  $\nu_0=1$ independently of
$n$. The RG problem is therefore well posed: it is sufficient to
impose the normalization condition eq.~(\ref{eq:polnorm}) to guarantee
that $\Psi$ has maximal rank.  

Numerical results are illustrated in Fig.~\ref{fig:bvmone} for both
symmetric and asymmetric cases: the two plots on top show the exact
solution $\cP_\text{sym}$ and its RG approximations for $N=1000$,
$Z_\pm=$ $Z\in\{16,32\}$, $\kappa=(2,2)$ and a bunch of values of
$n$; the central plots show results for the same physical setup, yet
with asymmetric choices of the weight index array $\kappa$; finally,
those at the bottom show the exact solution $\cP$ and its RG
approximations for $N=1000$, $Z_+=4$, $Z_-\in\{16,32\}$ and
$\kappa=(2,2)$. In Fig.~\ref{fig:bvmspectra}, we report plots of the
Dirichlet spectra obtained in the symmetric case with
$\kappa=(2,2)$. A few comments are in order: 
\begin{itemize}
\item{positiveness is violated at small values of $n$ for essentially
  all physical setups, but soon recovered at larger $n$;} 
\item{distributional convergence is reached at $n=2Nz-2$ in the
  symmetric case. It could not be otherwise: in order to exactly
  represent a polynomial $P$ with $\deg\{P\}=n$ by another polynomial
  $Q$, it must be $\deg\{Q\}\ge n$;} 
\item{convergence deteriorates in the symmetric case for $\kappa_1\ne
  \kappa_2$. Broadly speaking, the measure weight $\cD_\kappa$
  overlaps  with both $V_\alpha$ and $\cLFP\cdot \cD_\gamma$ in
  $\psi_{\alpha\gamma}$. If the mass of $\cP$ concentrates in a given
  subset of $\bar T_Q(s)$, it is recommendable not to use a weight
  function whose mass concentrates elsewhere. Although this suggestion
  is only useful once $\cP$ is known, symmetries should be taken into
  account in order to properly choose $\kappa$;} 
\item{the RG method works well also in the asymmetric case, where
  $\cP$ is a rational function. Here, exact convergence is expected to
  be reached only asymptotically;} 
\item{the Dirichlet spectra look rather localized. Instead of 
  ordering $c_{(\gamma_1,\gamma_2)}$ according to the RLO, in
  Fig.~\ref{fig:bvmspectra} we plot data against the relative 1--norm
  distance of $(\gamma_1,\gamma_2)$ from} 
\begin{equation}
  (\bar\gamma_1,\bar\gamma_2)=
  \underset{(\gamma_1,\gamma_2)\in\Omega_n}{\operatorname{argmax}}  
  \left\{c_{(\gamma_1,\gamma_2)}\right\} =
  \left(\frac{n}{2}+1,\frac{n}{2}+1\right)\,,\qquad n\text{
    even}\,. 
\end{equation}
\item[]{The outcome is evidently an exponential decrease for
  $n=2Nz-2$, with an  increasingly marked bending at larger values of
  $n$. An exponential behaviour is not surprising in consideration
  that $\cP_\text{sym}$ is essentially a  centered Gaussian
  distribution;}  
\item{the bending at $n>2Nz-2$ is clearly due to the
  non--orthogonality of $\{\cD_\gamma\}$.} 
\end{itemize}

\section{Example~2: multi--state voter model with zealots}

As a second case study for the RG method, we examine the multi--state
voter model with zealots, a generalization of the binary version
considered so far, where both dynamic agents and zealots carry an
opinion $v\in\{1,\ldots,Q\}$. The ordering dynamics of the model with
no zealots has been discussed in~\cite{Starnini}, while a variant with
committed agents on a weighted network has been more recently studied
in~\cite{Chen}. Here, we are interested in a simple formulation of the
model with $N$ agents on the complete graph, for which we expect  the
mean field description to work well. Let $N_k$ and $Z_k$
denote respectively the number of dynamic agents and zealots with
$v=k$. We  define $\phi_k = N_k/N$, $z_k=Z_k/N$ and
$s=1-\sum_{k=1}^{Q-1}z_k$. The FPE reads 
\begin{equation}
0 = -\sum_{\ell=1}^{Q-1}\partial_\ell [A_\ell(\phi)P(\phi)] +
\frac{1}{2}\sum_{\ell,m=1}^{Q-1}\partial_\ell\partial_m[B_{\ell
    m}(\phi)P(\phi)] 
\end{equation}
with drift and diffusion coefficients given by
\begin{align}
A_\ell(\bar\phi) & = z_\ell s - (1-s)\phi_\ell\,,\\[2.0ex]
B_{\ell m}(\bar\phi) & = \frac{\delta_{\ell
    m}}{N}[(\phi_\ell+z_\ell)(s-\phi_\ell) + \phi_\ell
  (1-z_\ell-\phi_\ell)] - \frac{1-\delta_{\ell m}}{N}[2\phi_\ell\phi_m
  + z_\ell\phi_m + z_m\phi_\ell]\,. 
\end{align}
To the best of our knowledge, no analytic solution of the FPE is
known in the literature. Therefore, the results of the RG method can
be only compared to numerical simulations.  Similar to the previous
section, the derivation of $\psi_{\alpha\gamma}$ requires a modest
algebraic effort. We have indeed 
\begin{align} 
& \langle
  V_\alpha,-\sum_{\ell=1}^{Q-1}\partial_\ell[A_\ell\cD_\gamma]\rangle_\kappa
  = (Q-1)(1-s)\langle V_\alpha,\cD_\gamma\rangle_\kappa +
  (1-s)\sum_{\ell=1}^{Q-1}\langle
  V_\alpha,\phi_\ell\partial_\ell\cD_\gamma\rangle_\kappa
  \nonumber\\[1.0ex] 
& \hskip 1.0cm -s\sum_{\ell=1}^{Q-1}z_\ell\langle
  V_\alpha,\partial_\ell \cD_\gamma\rangle_\kappa\,, 
\end{align}
\begin{align}
& \langle
  V_\alpha,\frac{1}{2}\sum_{m,\ell=1}^{Q-1}\partial_\ell\partial_m[B_{\ell
      m}\cD_\gamma]\rangle_\kappa = -\frac{Q(Q-1)}{N}\langle
  V_\alpha,\cD_\gamma\rangle_\kappa -
  \frac{2Q}{N}\sum_{\ell=1}^{Q-1}\langle
  V_\alpha,\phi_\ell\partial_\ell \cD_\gamma\rangle_\kappa
  \nonumber\\[1.0ex] 
& \hskip 1.0cm +\frac{1}{N}\sum_{\ell=1}^{Q-1}(1+s-Qz_\ell)\langle
  V_\alpha,\partial_\ell\cD_\gamma \rangle_\kappa -
  \frac{1}{N}\sum_{\ell=1}^{Q-1}\langle
  V_\alpha,\phi_\ell^2\partial_\ell^2\cD_\gamma \rangle_\kappa
  \nonumber
\end{align}
\begin{align}
& \hskip 1.0cm + \frac{1}{2N}\sum_{\ell=1}^{Q-1}(1+s-2z_\ell)\langle
  V_\alpha,\phi_\ell\partial_\ell^2\cD_\gamma \rangle_\kappa +
  \frac{s}{2N}\sum_{\ell=1}^{Q-1}z_\ell\langle V_\alpha,
  \partial_\ell^2\cD_\gamma \rangle_\kappa\nonumber\\[1.0ex] 
& \hskip 1.0cm -\frac{1}{2N}\sum_{\ell\ne m}^{1\ldots Q}\bigl[2\langle
    V_\alpha,\phi_\ell\phi_m\partial_\ell\partial_m\cD_\gamma
    \rangle_\kappa + z_\ell\langle
    V_\alpha,\phi_m\partial_\ell\partial_m\cD_\gamma\rangle_\kappa +
    z_m\langle
    V_\alpha,\phi_\ell\partial_\ell\partial_m\cD_\gamma\rangle_\kappa
    \bigr]\,, 
\end{align}
and we simply need to express the various scalar products in terms of
the matrix elements of $\chi$ via
eqs.~(\ref{eq:firstalg})--(\ref{eq:lastalg}). To give a feeling of the
goodness of the approximation, in Fig.~\ref{fig:msvm} we qualitatively
compare the histogram of the probability density obtained 
from Monte Carlo simulations of the model (top left) and the RG
approximation (top right) for a physical setup with $Q=3$, $N=1000$,
$Z_1=Z_2=Z_3=4$ and RG parameters $n=12$, $\nbc=20$ and
$\kappa=(2,2,2)$.  

It is interesting to examine how much $\cP_n$ depends upon the number
$\nbc$ of boundary conditions. 2--norm distances can be easily
evaluated once the RG coefficients are known. If $\cP_n^{(1)} =
\sum_{\gamma\in\Omega_n}c^{(1)}_\gamma\cD_\gamma$ and  $\cP_n^{(2)} =
\sum_{\gamma\in\Omega_n}c^{(2)}_\gamma\cD_\gamma$, then it can be
shown that 
\begin{equation}
||\cP_n^{(1)} - \cP_n^{(2)}||^2_2 = s^{1-Q}
\sum_{\gamma,\eta\in\Omega_n}\left[c^{(1)}_\gamma-c^{(2)}_\gamma\right]\left[c^{(1)}_\eta-c^{(2)}_\eta\right] 
\frac{\Gamma(|\gamma|)\Gamma(|\eta|)}{\Gamma(|\gamma|+|\eta|-Q)}\prod_{k=1}^{Q} 
\frac{\Gamma(\gamma_k+\eta_k-1)}{\Gamma(\gamma_k)\Gamma(\eta_k)}\,.
\end{equation}
\begin{figure}[t!]
  \centering
  \begin{minipage}[t]{0.45\textwidth}
    \includegraphics[width=0.95\textwidth]{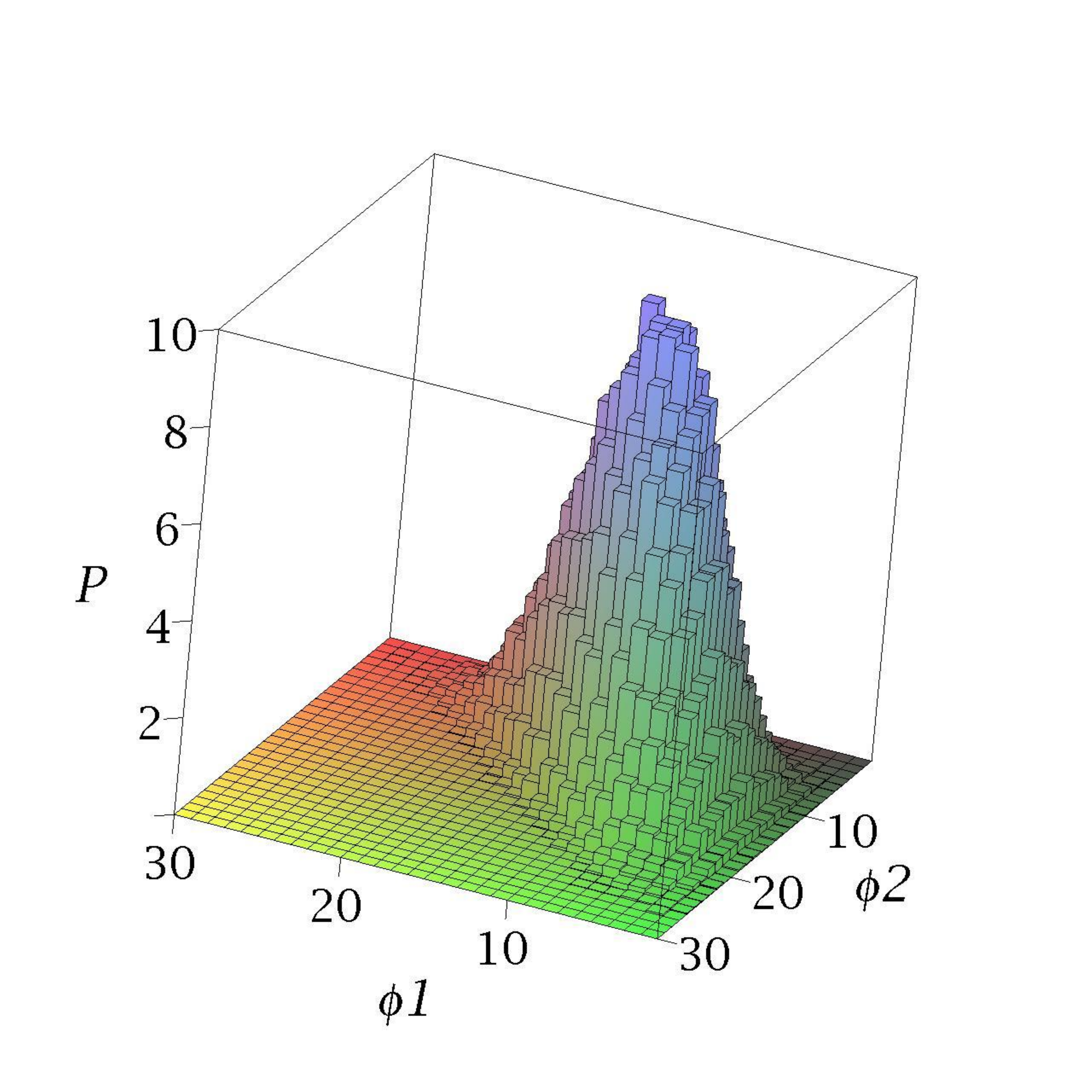}
  \end{minipage}
  \begin{minipage}[t]{0.45\textwidth}
    \includegraphics[width=0.95\textwidth]{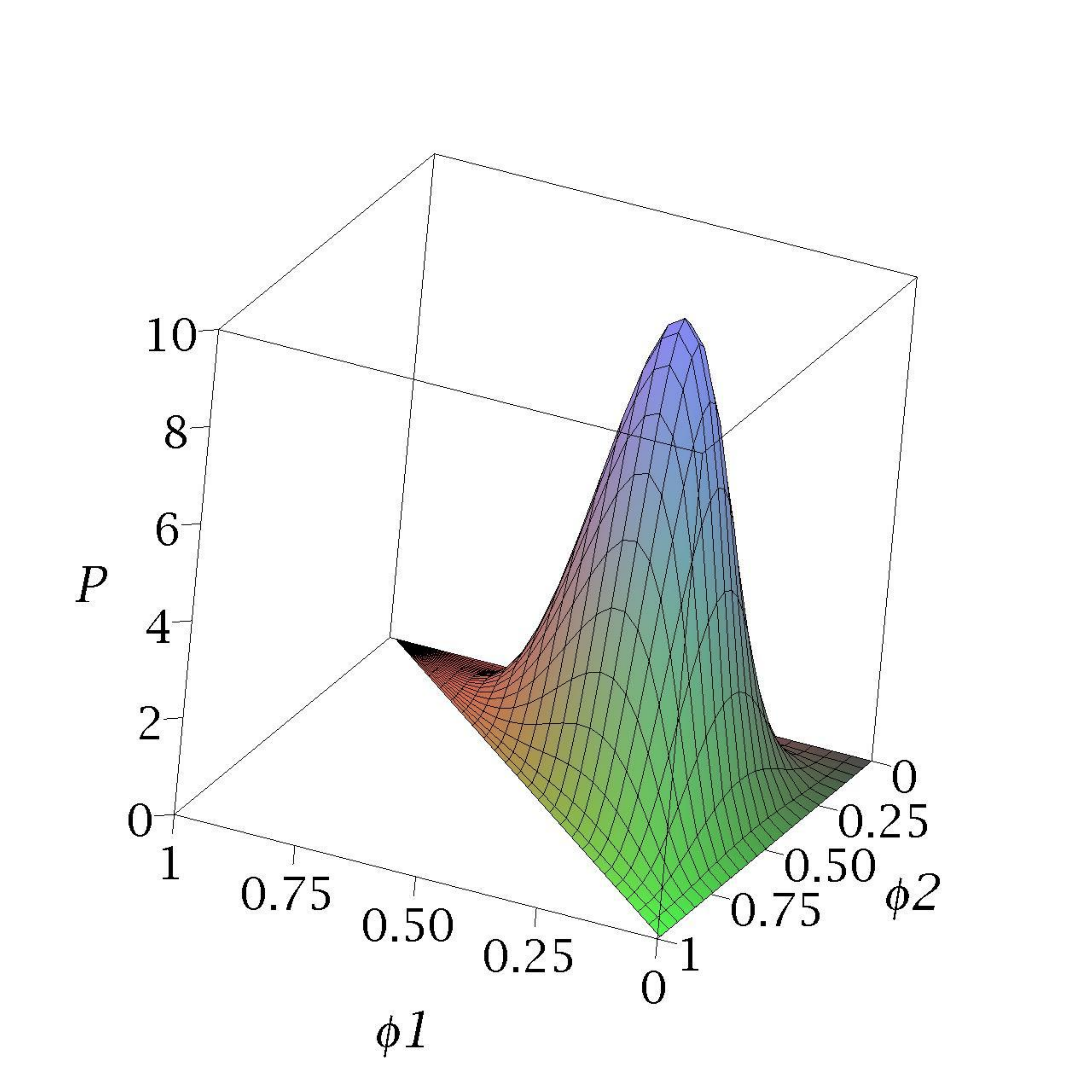}
  \end{minipage}
  \vskip 0.5cm
  \begin{minipage}[t]{0.45\textwidth}
    \hskip 0.5cm
    \includegraphics[width=0.95\textwidth]{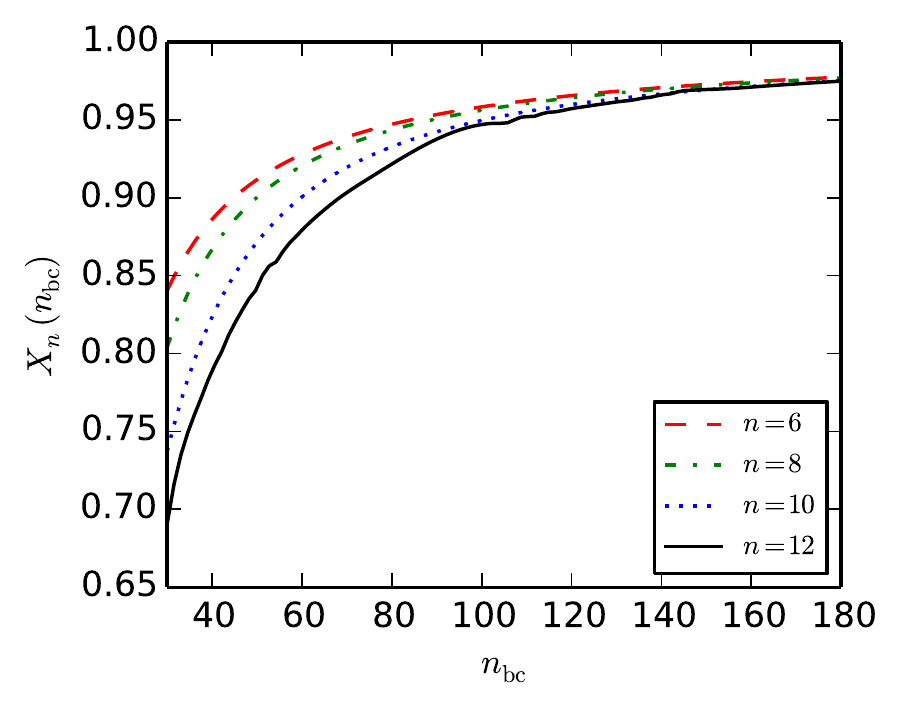}
  \end{minipage}
  \begin{minipage}[t]{0.45\textwidth}
  {
    \small
    \hskip 2.0cm
    \begin{tabular}{r|c}
      \\[-6.4cm]
      \hline\hline\\[-2.0ex]
      $n$ & $\log_{10}\text{cond}\{\trans{\Psi}\cdot\Psi\}$ \\[0.2ex]
      \hline\\[-2.3ex]
      6 & 10.67(1) \\
      8 & 14.85(1) \\
      10 & 19.17(1) \\
      12 & 23.69(1) \\
      14 & 28.24(1) \\
      16 & 32.73(1) \\
      18 & 37.16(1) \\[0.6ex]
      \hline\hline
    \end{tabular}
  }
  \end{minipage}
  \caption{ \footnotesize (top left) Probability density of the
    multi--state voter model from Monte Carlo simulations (axes have
    been rescaled so as to host a 30--bins histogram with unitary bin    
    size). (top right) RG probability density with parameters $n=12$,    
    $\nbc=20$ and
    $\kappa=(2,2,2)$. (bottom left) Sensitivity of $\cP_n$ to the
    number $\nbc$ of boundary conditions, see
    eq.~(\ref{eq:Xn}). (bottom right) condition number of
    $\trans{\Psi}\cdot\Psi$. All plots refer to a physical setup with
    $Q=3$, $N=1000$, $Z_1=Z_2=Z_3=4$.} 
\label{fig:msvm}
\end{figure}
\noindent In order to assess the sensitivity of $\cP_n$ to $\nbc$, we
could look at $||\cP_n|_{\nbc+2}-\cP_n|_{\nbc}||_2^2$ as a function of
$\nbc$. Unfortunately, this quantity depends strongly on the polynomial
degree $n$, hence it becomes difficult to compare distances
corresponding to different values of $n$. A smoother behaviour is
displayed by the distance ratio 
\begin{equation}
X_n(\nbc) =
\sqrt{\frac{||\cP_n|_{\nbc+2}-\cP_n|_{\nbc}||_2^2}{||\cP_n|_{\nbc}-\cP_n|_{\nbc-2}||_2^2}}\qquad
\left(\text{notice that} \lim_{\nbc\to\infty}X_n(\nbc) = 1 \right)\,, 
\label{eq:Xn}
\end{equation}
which we plot against $\nbc$ in Fig.~5 (bottom left), once more for
$Q=3$, $N=1000$, $Z_1=Z_2=Z_3=4$ and $\kappa=(2,2,2)$. For $n>12$,
limits of our computer implementation emerge: $X_n(\nbc)$ becomes
numerically unstable due to large cancellations occurring when
subtracting the coefficients $c_\gamma$, hence we give up reporting
it. Anyway, the plot shows that solutions of higher degree are more
sensitive to the number of boundary conditions. This looks natural if
one considers that the larger $n$ the more $\cP_n$ fluctuates on the
boundary hypersurfaces: in order to {\it gentle} the orthogonal
probability flux crossing the boundary, this must be forced to 
vanish at more and more boundary points. By construction
$X_n$ carries no information about the overall scale of the 2--norm
distances. This turns out to be very small for all $n$ and
$\nbc\gtrsim 10$ (this estimate is likely to increase for larger values of $Q$).

It is likewise interesting to look at the condition number of
$\trans{\Psi}\cdot\Psi$. As the table in Fig.~5 (bottom right)
shows, this blows up exponentially as $n$
increases, while it is rather insensitive to $\nbc$ (the uncertainty reported
in the table measures the variation range for $20\le\nbc\le
200$). The exponential enhancement with $n$ requires a robust
algorithm in order to perform the matrix inversion, as already
observed in sect.~3.   

Finally, the above discussion concerns only the analytic properties of
$\cP_n$. In order to make a quantitative comparison between $\cP_n$
and the empirical probability density obtained from the  Monte Carlo
(MC) simulation of the multi--state voter model, we can look at the respective
distributional moments. Those of $\cP_n$ can be easily worked--out and
exactly expressed as functions of the coefficients $\{c_\gamma\}$. In
particular, the first two moments are given by   
\begin{align}
& \E\left[\phi_k\, |\, \cP_n\right]  =
s\sum_{\gamma\in\Omega_n}c_\gamma\frac{\gamma_k}{|\gamma|}\,,\\[1.0ex] 
& \E\left[\phi_k^2\, |\, \cP_n\right] =
s^2\sum_{\gamma\in\Omega_n}c_\gamma\frac{\gamma_k(\gamma_k+1)}{|\gamma|(|\gamma|+1)}\,,
\qquad \E\left[\phi_j\phi_k\, |\, \cP_n\right]_{j\ne k} =
s^2\sum_{\gamma\in\Omega_n}c_\gamma\frac{\gamma_j\gamma_k}{|\gamma|(|\gamma|+1)}\,. 
\end{align}
Numerical estimates look rather stable against changes of $n$ and
$\nbc$. For physical parameters as above all RG approximations give
 $\E[\phi_k]_\text{RG} = 0.32933\ldots = s/3$. However, the first
moment is not indicative, as it just results from the symmetry of the
setup. In order to make a real comparison, we have to look at the second
moments. Our best estimates from RG approximations are
$\E[\phi_k^2]_\text{RG} = 0.1253\ldots$ to be compared to
$\E[\phi_k^2]_\text{MC} = 0.124(1)$ and for $j\ne k$, $\E[\phi_j\phi_k]_\text{RG} =
0.1000\ldots$ to be compared to
$\E[\phi_j\phi_k]_\text{MC}=0.1004(5)$. As can be seen, results are in
very good agreement.  

\section{Symmetry considerations}

The symmetry group of the simplex is the symmetric group. Since the
implementation of the RG method becomes numerically demanding at
large $Q$, it is worthwhile discussing if and how permutational
symmetries can help reduce the computational work load.   

\subsection{Permutational symmetry of the coefficients $\chi_{\alpha\gamma}$}

We first observe that the choice of the weight index array $\kappa$ is
totally arbitrary, yet different values of it correspond to different
orthogonal bases. A convenient option is the isotropic one, namely 
\begin{equation}
\kI = (\underbrace{\hat \kappa,\ldots ,\hat \kappa}_{Q-1\text{ times}},\bar\kappa)\,, 
\end{equation}
which for any $Q$ depends only on two integer values $\hat\kappa$ and
$\bar\kappa$. If we denote by $S_{Q-1}$ the set of permutations of
$\{1,\ldots,Q-1\}>$ and for
$\sigma\in S_{Q-1}$ we define $\sigma\cdot\bar\phi \equiv
(\phi_{\sigma(1)},\ldots,\phi_{\sigma(Q-1)})$, $\sigma\cdot\alpha =
(\alpha_{\sigma(1)},\ldots,\alpha_{\sigma(Q-1)})$ for $\alpha\in
\dN_0^{Q-1}$ and $\sigma\cdot\gamma
= (\gamma_{\sigma(1)},\ldots,\ldots,\gamma_{\sigma(Q-1)},\gamma_Q)$
for $\gamma\in\dN^Q$, then we immediately see that
$\cD_\kI(\sigma\cdot\bar\phi) = \cD_\kI(\bar\phi)$. For any other
choice of the index array, it holds
\begin{align}
\cD_\gamma(\sigma\cdot\bar\phi) & = \frac{\Gamma(|\gamma|)}{\prod_{m=1}^Q\Gamma(\gamma_m)}s^{1-|\gamma|}\left[\prod_{m=1}^{Q-1}\phi_{\sigma(m)}^{\slashed{\gamma}_m}\right](s-|\bar\phi|)^{\slashed{\gamma_Q}} \nonumber\\[2.0ex]
& = \frac{\Gamma(|\gamma|)}{\prod_{m=1}^Q\Gamma(\gamma_m)}s^{1-|\gamma|}\left[\prod_{m=1}^{Q-1}\phi_m^{\slashed{\gamma}_{\sigma^{-1}(m)}}\right](s-|\bar\phi|)^{\slashed{\gamma}_Q} = \cD_{\sigma^{-1}\cdot\gamma}(\bar\phi)\,,
\label{eq:dircov}
\end{align}
with $\sigma^{-1}$ denoting the inverse permutation of
$\sigma$. Remarkably, a property analogous to eq.~(\ref{eq:dircov}) is
also fulfilled by the orthogonal polynomials provided $\kappa=\kI$, namely
\begin{myprop}
If $\kappa= \kI$, then $V_\alpha(\sigma\cdot\bar\phi) = V_{\sigma^{-1}\cdot\alpha}(\bar\phi)$, $U_\alpha(\sigma\cdot\bar\phi) = U_{\sigma^{-1}\cdot\alpha}(\bar\phi)$ and  $f_{\sigma\cdot\alpha} = f_\alpha$ for any $\alpha\in\dN_0^{Q-1}$, $\sigma\in S_{Q-1}$ and $\bar\phi\in \bar T_Q(s)$. 
\end{myprop}
\noindent\emph{Proof}. With regard to $V_\alpha$, we first notice by
direct inspection that 
\begin{equation}
v_{\alpha(\sigma\cdot\beta)}(\kappa) = v_{(\sigma^{-1}\cdot{\alpha})\beta}(\sigma^{-1}\cdot\kappa)\,.
\label{eq:vsymm}
\end{equation}
Therefore, we have
\begin{align}
V_\alpha\left(\sigma\cdot \bar\phi\right) & =
\sum_{\beta\le\alpha}v_{\alpha\beta}(\kappa)(\sigma\cdot \bar\phi)^\beta =  \sum_{\beta\le\alpha}v_{\alpha\beta}(\kappa)\bar\phi^{\sigma^{-1}\cdot\beta} = \sum_{\sigma\cdot\beta\le\alpha}v_{\alpha(\sigma\cdot\beta)}(\kappa)\bar\phi^\beta \nonumber\\[2.0ex]
& = \sum_{\sigma\cdot \beta\le\alpha}v_{\sigma^{-1}\cdot\alpha\beta}(\sigma^{-1}\cdot\kappa)\bar\phi^\beta\,.
\end{align}
Moreover,
\begin{align}
\sum_{\sigma\cdot\beta\le\alpha} =
\sum_{\beta_{\sigma(1)}=0}^{\alpha_1}\ldots
\sum_{\beta_{\sigma(d)}=0}^{\alpha_d} =
\sum_{\beta_1=0}^{\alpha_{\sigma^{-1}(1)}}\ldots
\sum_{\beta_d=0}^{\alpha_{\sigma^{-1}(d)}} = \sum_{\beta\le \sigma^{-1}\cdot \alpha}
\end{align}
whence we conclude
\begin{equation}
V_\alpha\left(\sigma\cdot\bar\phi\right) = \sum_{\beta\le\sigma^{-1}\cdot \alpha}v_{(\sigma^{-1}\cdot\alpha)\beta}(\sigma\cdot\kappa)\bar\phi^\beta = V_{\sigma^{-1}\cdot\alpha}(\bar\phi)\quad\text{if}\quad \sigma\cdot\kappa = \kappa\,.
\end{equation}
Analogously, we have
\begin{align}
U_\alpha(\sigma\cdot\bar\phi) & = \frac{1}{\cD_\kappa(\sigma\cdot\bar\phi)}\frac{\partial^{|\alpha|}}{\partial\phi_{\sigma(1)}^{\alpha_1}\ldots\partial\phi_{\sigma(d)}^{\alpha_d}}\left\{\prod_{m=1}^{Q-1}\phi_{\sigma(m)}^{\alpha_m+\kappa_m-1}(1-|\bar\phi|)^{|\alpha|+\kappa_{Q}-1}\right\} \nonumber\\[2.0ex]
& = \frac{1}{\cD_{\sigma^{-1}\cdot\kappa}(\bar\phi)}\partial^{|\alpha|}_{\sigma^{-1}\cdot\alpha}\left\{\prod_{m=1}^{Q-1}\phi_m^{\alpha_{\sigma^{-1}(m)}+\kappa_{\sigma^{-1}(m)}-1}(1-|\bar\phi|)^{|\alpha|+\kappa_{Q}-1}\right\}\nonumber\\[2.0ex] & = U_{\sigma^{-1}\cdot \alpha}(\bar\phi)\quad\text{if}\quad\sigma^{-1}\cdot\kappa = \kappa\,.
\end{align}
It is trivially clear that $f_{\sigma\cdot\alpha}=f_\alpha$ if $\kappa
= \kI$.\qed 

\begin{figure}[t!]
\small
  \vskip 0.3cm
  \centering
  \colorbox{myblue}
  {
    \parbox{0.4\textwidth}
           {
             \vbox{
               \texttt{for}\ k \texttt{from } $0$ \texttt{ to } $n$                                          \texttt{                               do}\\[0.5ex]
\hbox{\texttt{\ \ for } $\alpha$ \texttt{\,in                   }$\fP_{k,Q-1}$\texttt{ do } }\\[0.5ex]
\hbox{\texttt{\ \ \ \ for } $\gamma$ \texttt{\,in               }$\Omega_n$\texttt{ do }} \\[0.5ex]
\hbox{\texttt{\ \ \ \ \ \ compute } 
$\chi_{\alpha\gamma} = \langle V_\alpha,\cD_\gamma\rangle_{\kI}$} \\[0.5ex]
\hbox{\texttt{\ \ \ \ \ \ for } $\beta$ \texttt{\,in } $\Pi(\alpha)/\{\alpha\}$\texttt{ do }} \\[0.5ex]
\hbox{\texttt{\ \ \ \ \ \ \ \ find } $\sigma\in S_{Q-1}: \sigma\cdot\beta = \alpha$}  \\[0.5ex]
\hbox{\texttt{\ \ \ \ \ \ \ \ assign } $\chi_{\beta(\sigma^{-1}\cdot\gamma)} \leftarrow \chi_{\alpha\gamma}$} \\[0.5ex]
\hbox{ \texttt{\ \ \ \ \ \ end do } }\\[0.5ex]
\hbox{ \texttt{\ \ \ \ end do } }\\[0.5ex]
\hbox{ \texttt{\ \ end do } }\\[0.5ex]
\hbox{ \texttt{end do } }\\[-3.0ex]
             }
           }
  }
  \vskip 0.2cm
  \caption{A convenient recipe to compute $\chi$ when $\kappa=\kI$.}
  \label{fig:chialg}
\end{figure}

\noindent From Prop.~3 it follows
\begin{align} 
\chi_{\alpha\gamma} & = \int_{\bar T_Q(s)}\rd\bar\phi\, V_\alpha(\bar\phi)\cD_\gamma(\bar\phi)\cD_\kI(\bar\phi) \!\!\!\! \stackrel {\phantom{\int_{a_a}} \bar\phi \to \sigma\cdot\bar\phi'\ \ }{=} \!\!\! \int_{\bar T_Q(s)}\rd\bar\phi'\, V_\alpha(\sigma\cdot \bar\phi')\cD_\gamma(\sigma\cdot\bar\phi')\cD_\kI(\sigma\cdot\bar\phi')\nonumber\\[2.0ex]
& = \ \int_{\bar T_Q(s)}\rd\bar\phi'\, V_{\sigma^{-1}\cdot\alpha}(\bar\phi') \cD_{\sigma^{-1}\cdot \gamma}(\bar\phi')\cD_\kI(\bar\phi') = \chi_{(\sigma^{-1}\cdot\alpha)( \sigma^{-1}\cdot\gamma)}\,\qquad \text{if} \ \kappa = \kI\,.
\end{align}
Accordingly, many matrix elements of $\chi$ are exactly the same,
which explains the little--square structure of Fig.~\ref{fig:chi}. In
order to establish a convenient way of computing $\chi$,  we introduce
the partition set 
\begin{equation} 
\fP_{n,Q-1} = \{\alpha\in \dN_0^{Q-1}: |\alpha|=n \ \text{ and }
\ \alpha_1\ge\alpha_2\ge\ldots\ge\alpha_{Q-1}\}\,, 
\label{eq:partone}
\end{equation}
and for $\alpha\in \dN_0^{Q-1}$ the permutation set
\begin{equation}
\Pi(\alpha) = \left\{\eta\in\dN_0^{Q-1}: \ \ \eta=\sigma\cdot\alpha
\text{ \ \ for some \ }\sigma\in S_{Q-1}\right\}\,. 
\label{eq:permutone}
\end{equation}
Obviously, if $\alpha$ has $m_1$ components equal to $a_1$, \ldots,
$m_r$ components equal to $a_r$, such that 
\begin{equation}
m_1 + \ldots + m_r = Q-1\,,\qquad a_1m_1+\ldots+a_rm_r = |\alpha|\,,
\end{equation}
then it holds
\begin{equation}
|\Pi(\alpha)| = \frac{(Q-1)!}{m_1!\ldots m_r!} = {Q-1\choose m_1,\ldots,m_r}\,.
\end{equation}
Partitions and permutations allow to decompose the index space of the
orthogonal polynomials as a union of disjoint sets, namely 
\begin{equation}
\{\alpha\in\dN_0^{Q-1}: |\alpha|\le n\} = \bigcup_{k=0}^n
\bigcup_{\alpha\in\fP_{k,Q-1}}\Pi(\alpha)\,. 
\label{eq:domdecomp}
\end{equation}
In Fig.~\ref{fig:chialg} we provide a recipe to compute $\chi$, which
is based on the above set decomposition and works correctly since
$\Omega_n$ is permutationally closed. Moreover, it is well known since
Euler's age \cite{knuth4a} that  $p_{n,Q-1} \equiv
|\mathfrak{P}_{n,Q-1}|$ can be obtained from the generating function 
\begin{equation}
f(x) = \prod_{n=1}^{Q-1}\frac{1}{1-x^k} = \sum_{n=0}^\infty p_{n,Q-1}x^k\,.
\end{equation}
If we define the truncated Taylor expansion
\begin{equation}
F_n(x) = \sum_{k=0}^np_{k,Q-1}x^k\,,
\end{equation}
then $q_{n,Q-1} = F_n(1)$ represents the total number of matrix rows
$\alpha$ for which $\chi_{\alpha\gamma}$ really needs to be
computed. In Table~\ref{tab:qkd}, we report $q_{n,Q-1}$ for the first
few values of $n$ and $Q$.  

\begin{table}[!t]
\small
\centering
  \begin{tabular}{c|cccccccccccccc}
    \hline\hline\\[-2.0ex]
    $n\backslash Q$ &  2 & 3 & 4 & 5 & 6 & 7 & 8 & 9 & 10 & 11 & 12 & 13
    & 14 & 15 \\[0.2ex] 
    \hline\\[-2.3ex]
0  & 1 & 1 & 1 & 1 & 1 & 1 & 1 & 1 & 1 & 1 & 1 & 1 & 1 & 1 \\
1  & 2 & 2 & 2 & 2 & 2 & 2 & 2 & 2 & 2 & 2 & 2 & 2 & 2 & 2 \\
2  & 4 & 4 & 4 & 4 & 4 & 4 & 4 & 4 & 4 & 4 & 4 & 4 & 4 & 4 \\
3  & 6 & 7 & 7 & 7 & 7 & 7 & 7 & 7 & 7 & 7 & 7 & 7 & 7 & 7 \\
4  & 9 & 11 & 12 & 12 & 12 & 12 & 12 & 12 & 12 & 12 & 12 & 12 & 12 & 12 \\
5  & 12 & 16 & 18 & 19 & 19 & 19 & 19 & 19 & 19 & 19 & 19 & 19 & 19 & 19 \\
6  & 16 & 23 & 27 & 29 & 30 & 30 & 30 & 30 & 30 & 30 & 30 & 30 & 30 & 30 \\ 
7  & 20 & 31 & 38 & 42 & 44 & 45 & 45 & 45 & 45 & 45 & 45 & 45 & 45 & 45 \\
8  & 25 & 41 & 53 & 60 & 64 & 66 & 67 & 67 & 67 & 67 & 67 & 67 & 67 & 67 \\ 
9  & 30 & 53 & 71 & 83 & 90 & 94 & 96 & 97 & 97 & 97 & 97 & 97 & 97 & 97 \\
10 & 36 & 67 & 94 & 113 & 125 & 132 & 136 & 138 & 139 & 139 & 139 & 139 & 139 & 139 \\
11  & 42 & 83 & 121 & 150 & 169 & 181 & 188 & 192 & 194 & 195 & 195 & 195 & 195 & 195 \\
12  & 49 & 102 & 155 & 197 & 227 & 246 & 258 & 265 & 269 & 271 & 272 & 272 & 272 & 272 \\ 
13  & 56 & 123 & 194 & 254 & 298 & 328 & 347 & 359 & 366 & 370 & 372 & 373 & 373 & 373 \\ 
14  & 64 & 147 & 241 & 324 & 388 & 433 & 463 & 482 & 494 & 501 & 505 & 507 & 508 & 508 \\ 
15  & 72 & 174 & 295 & 408 & 498 & 564 & 609 & 639 & 658 & 670 & 677 & 681 & 683 & 684 \\[2.0ex]
\hline\hline
  \end{tabular}
  \caption{First coefficients $q_{n,Q-1}$.}
  \label{tab:qkd}
\end{table}

\subsection{Symmetric solutions of the Fokker--Planck equation}

If the FPE is symmetric under a subset of index permutations of
$\bar\phi$, then its solution is expected to have the same symmetry
(we assume here that no spontaneous symmetry breaking occurs). A
permutationally symmetric FPE describes a system where no
physical state is \emph{a priori} favoured with
respect to the others, a frequent case in phenomenological applications. 
For simplicity's sake, we assume that the FPE is maximally
symmetric, \ie it is symmetric under
$\bar\phi\to\sigma\cdot\bar\phi$,  $\forall \sigma\in S_{Q-1}$. As an
example, the reader could consider a multi--state voter model with
zealots, where $Z_k=Z$ for $k=1,\ldots,Q$ and the network topology
preserves the symmetry (for a non--trivial instance, see ref.~\cite{palombi}). 
Imposing that $\cP_n$ is permutationally invariant yields
\begin{align}
\cP_n(\sigma\cdot\bar\phi) & =\sum_{\gamma\in\Omega_n}c_\gamma\cD_\gamma(\sigma\cdot\bar\phi) =\sum_{\gamma\in\Omega_n}c_\gamma\cD_{\sigma^{-1}\cdot\gamma}(\bar\phi)\nonumber\\[1.0ex] & =\sum_{\gamma\in\Omega_n}c_{\sigma\cdot\gamma}\cD_{\gamma}(\bar\phi)
=\sum_{\gamma\in\Omega_n}c_{\gamma}\cD_{\gamma}(\bar\phi) = \cP_n(\bar\phi)\,,
\end{align}
and since $\{\cD_\gamma\}$ is a polynomial basis, we infer
$c_{\sigma\cdot\gamma}=c_\gamma$,
$\forall \gamma\in\Omega_n$ and  $\forall \sigma\in S_{Q-1}$. Of
course, this result can be fruitfully used as a check of the numerical
implementation of the RG method. Nevertheless, in this section we
would like to discuss whether we can include such information even in 
the theoretical construction of the weak solution. In analogy with
eqs.~(\ref{eq:partone})--(\ref{eq:permutone}), we introduce the
partition set
\begin{equation} 
\fL_n = \{\gamma\in \Omega_n: \ \gamma_1\ge\gamma_2\ge\ldots\ge\gamma_{Q-1}\}\,,
\label{eq:parttwo}
\end{equation}
and for $\gamma\in\Omega_n$ the permutation set
\begin{equation}
 \Lambda(\gamma) = \left\{\eta\in\Omega_n: \ \ \eta=\sigma\cdot\gamma
 \text{ \ \ for some \ }\sigma\in S_{Q-1}\right\}\,. 
 \label{eq:permuttwo}
\end{equation}
If $\cP_n$ is permutationally symmetric, then it can be written as
\begin{equation}
\cP_n(\bar\phi) =
\sum_{\gamma\in\fL_n}c_\gamma\sum_{\eta\in\Lambda(\gamma)}\cD_\eta(\bar\phi)
= \sum_{\gamma\in\fL_n}\hat c_\gamma
\frac{1}{|\Lambda(\gamma)|}\sum_{\eta\in\Lambda(\gamma)}\cD_\eta(\bar\phi)
= \sum_{\gamma\in\fL_n}\hat c_\gamma \,\cS\cdot\cD_\gamma(\bar\phi)\,,
\label{eq:Psym}
\end{equation}
where we have introduced the rescaled coefficient $\hat c_\gamma =
|\Lambda(\gamma)|c_\gamma$ and 
the symmetrized Dirichlet distribution $\cS\cdot\cD_\gamma =
|\Lambda(\gamma)|^{-1}\sum_{\eta\in\Lambda(\gamma)}\cD_\eta$. Eq.~(\ref{eq:Psym})
tells us that a symmetric RG solution can be represented as a linear
combination of symmetrized Dirichlet distributions (which are
symmetric!). Can we reformulate the whole RG method so as to only make 
use of symmetrized Dirichlet distributions? The answer is affirmative, yet the
reader should not undervalue technicalities. 

\noindent\emph{i}) As a preliminary observation, we argue that a
symmetrized Dirichlet distribution faithfully decomposes into a basis
of symmetrized orthogonal polynomials. To this aim, we first need to
examine the permutational properties of the coefficients
$\{d_{\gamma\beta}\}$ connecting the Dirichlet basis $\{\cD_\gamma\}$
to the Appel basis $\{U_\beta\}$, see eq.~(\ref{eq:Dirproj}). Under
the isotropic assumption, a permutation of the components of
$\bar\phi$ results in  
\begin{align}
\cD_\gamma(\sigma\cdot\bar\phi) &
=\sum_{|\beta|\le|\slashed{\gamma}|}d_{\gamma\beta}\,U_\beta(\sigma\cdot\bar\phi)
=\sum_{|\beta|\le|\slashed{\gamma}|}d_{\gamma\beta}\,U_{\sigma^{-1}\cdot\beta}(\bar\phi)
=\sum_{|\beta|\le|\slashed{\gamma}|}d_{\gamma(\sigma\cdot\beta)}\,U_{\beta}(\bar\phi)\,. 
\end{align}
However, 
\begin{equation}
\cD_\gamma(\sigma\cdot\bar\phi) =
\cD_{\sigma^{-1}\cdot\gamma}(\bar\phi) \ =
\sum_{|\beta|\le|\slashed{\gamma}|}d_{(\sigma^{-1}\cdot\gamma)\beta}\,U_\beta(\bar\phi)\,, 
\end{equation}
whence we infer
\begin{equation}
 \sum_{|\beta|\le|\slashed{\gamma}|}d_{(\sigma^{-1}\cdot\gamma)\beta}\,U_\beta(\bar\phi)
\ =\sum_{|\beta|\le|\slashed{\gamma}|}d_{\gamma(\sigma\cdot\beta)}\,U_{\beta}(\bar\phi)\,.
\end{equation}
Since $\{U_\beta\}$ is a polynomial basis, we conclude that
$d_{\gamma(\sigma\cdot\beta)} =
d_{(\sigma^{-1}\cdot\gamma)\beta}$ (of course, this could have
been equivalently obtained via the identity $\chi_{\alpha\gamma} =
\chi_{(\sigma^{-1}\cdot\alpha)(\sigma^{-1}\cdot\gamma)}$, discussed in
sect.~6.1). Then, we apply the symmetrization
operator $\cS$ to both sides of eq.~(\ref{eq:Dirproj}), namely
\begin{align}
\cS\cdot\cD_\gamma(\bar\phi) & =
\frac{1}{|\Lambda(\gamma)|}\,\sum_{\eta\in\Lambda(\gamma)}\sum_{|\beta|\le
  |\slashed{\gamma}|} d_{\eta\beta}
U_\beta(\bar\phi)\nonumber\\[2.0ex] 
& = \sum_{|\beta|\le
  |\slashed{\gamma}|}\left[\frac{1}{|\Lambda(\gamma)|}\,\sum_{\eta\in\Lambda(\gamma)}d_{\eta\beta}\right] 
U_\beta(\bar\phi) =  \sum_{|\beta|\le |\slashed{\gamma}|}
e_{\gamma\beta}U_\beta(\bar\phi)\,, 
\end{align}
with $e_{\gamma\beta}\equiv
|\Lambda(\gamma)|^{-1}\,\sum_{\eta\in\Lambda(\gamma)}d_{\eta\beta} =
\cS\cdot d_{\gamma\beta}$. We can easily work out the permutational
properties of the coefficients $\{e_{\gamma\beta}\}$. We have indeed
\begin{align}
e_{\gamma(\sigma\cdot\beta)}
=\frac{1}{|\Lambda(\gamma)|}\sum_{\eta\in\Lambda(\gamma)}d_{\eta(\sigma\cdot\beta)}
=
\frac{1}{|\Lambda(\gamma)|}\sum_{\eta\in\Lambda(\gamma)}d_{(\sigma^{-1}\cdot\eta)\beta}
= e_{\gamma\beta}\,, 
\end{align}
as $\Lambda(\gamma) = \Lambda(\sigma\cdot\gamma)$ for any $\sigma\in S_{Q-1}$. For the same reason, it holds
\begin{equation}
e_{(\sigma\cdot\gamma)\beta}
=\frac{1}{|\Lambda(\sigma\cdot\gamma)|}\sum_{\eta\in\Lambda(\sigma\cdot\gamma)}d_{\eta\beta}
=
\frac{1}{|\Lambda(\gamma)|}\sum_{\eta\in\Lambda(\gamma)}d_{\eta\beta}
= e_{\gamma\beta}\,. 
\end{equation}
Therefore, we conclude that $e_{\gamma(\sigma\cdot\beta)} =
e_{(\sigma\cdot\gamma)\beta} = e_{\gamma\beta}$. By decomposing
the set $\{|\beta|\le|\slashed{\gamma}|\}$ according to
eq.~(\ref{eq:domdecomp}), we finally obtain
\begin{align}
\cS\cdot\cD_\gamma(\bar\phi) & =
\sum_{k=0}^{|\slashed{\gamma}|}\sum_{\beta\in\fP_{k,Q-1}}\sum_{\alpha\in\Pi(\beta)}e_{\gamma\alpha}\,U_\alpha(\bar\phi) 
 =
 \sum_{k=0}^{|\slashed{\gamma}|}\sum_{\beta\in\fP_{k,Q-1}}e_{\gamma\beta}\sum_{\alpha\in\Pi(\beta)}\,U_\alpha(\bar\phi) 
 \nonumber\\[2.0ex]
& = \sum_{k=0}^{|\slashed{\gamma}|}\sum_{\beta\in\fP_{k,Q-1}}\hat e_{\gamma\beta}\,\cS\cdot U_\beta(\bar\phi)\,,
\label{eq:symdecomp}
\end{align}
with $\hat e_{\gamma\beta} \equiv
|\Pi(\beta)|e_{\gamma\beta}$. Eq.~(\ref{eq:symdecomp}) tells us that
since $\cS\cdot\cD_\gamma$ is a symmetric function, it
decomposes faithfully into a set of symmetric polynomials $\{\cS\cdot
U_\beta\}$. In addition, we notice that if 
$\alpha,\beta\in\dN_0^{Q-1}$, then either
$\Pi(\alpha)\cap\Pi(\beta)=\emptyset$ or $\Pi(\alpha)=\Pi(\beta)$. 
Therefore, it makes sense to define
\begin{equation}
\delta_{\Pi(\beta)\Pi(\gamma)} = \left\{\begin{array}{ll} 1 & \text{
  if }\ \Pi(\beta) = \Pi(\gamma)\,,\\[1.0ex] 0 & \text{ if
}\ \Pi(\beta)\cap\Pi(\gamma) = \emptyset\,.\end{array}\right. 
\end{equation}
Finally, we observe that
\begin{align}
\langle\cS\cdot V_\alpha,\cS\cdot U_\beta\rangle_\kappa &
=\frac{1}{|\Pi(\alpha)}\frac{1}{|\Pi(\beta)|}\sum_{\epsilon\in\Pi(\alpha)}\sum_{\eta\in\Pi(\beta)}
\langle V_\epsilon,U_\eta\rangle_\kappa \nonumber\\[2.0ex] 
&  =
\frac{1}{|\Pi(\alpha)}\frac{1}{|\Pi(\beta)|}\sum_{\epsilon\in\Pi(\alpha)}\sum_{\eta\in\Pi(\beta)}
f_\epsilon \delta_{\epsilon\eta} =
\frac{f_\alpha}{|\Pi(\alpha)|}\delta_{\Pi(\alpha)\Pi(\beta)}\,, 
\end{align}
since $f_\alpha$ is permutationally invariant under the isotropic
assumption. Now, projecting $\cS\cdot\cD_\gamma$ onto $\cS\cdot
V_\alpha$ with  $|\alpha|\le|\slashed{\gamma}|$ yields 
\begin{align}
\langle \cS\cdot V_\alpha,\cS\cdot\cD_\gamma\rangle_\kappa &
=\sum_{k=0}^{|\slashed{\gamma}|}\sum_{\beta\in\fP_{k,Q-1}}\hat
e_{\gamma\beta}\,\langle \cS V_\alpha,\cS U_\beta\rangle_\kappa 
 \nonumber\\[1.0ex]
& = \sum_{k=0}^{|\slashed{\gamma}|}\sum_{\beta\in\fP_{k,Q-1}}
 \frac{\hat
   e_{\gamma\beta}f_\alpha}{|\Pi(\alpha)|}\delta_{\Pi(\alpha)\Pi(\beta)}
 = e_{\gamma\alpha}f_\alpha\,, 
\end{align}
while the above scalar product vanishes for
$|\alpha|>|\slashed{\gamma}|$. We conclude 
\begin{equation}
\langle \cS\cdot V_\alpha,\cS\cdot\cD_\gamma\rangle_\kappa =
\left\{\begin{array}{ll} \chi_{\alpha\gamma}\cdot\overleftarrow{\cS} & \text{ if
}\ |\alpha|\le|\slashed{\gamma}|\,,\\[2.0ex] 0 & \text{
  otherwise\,,}\end{array}\right.
\label{eq:symscalprod}
\end{equation}
where $\overleftarrow{\cS}$ symmetrizes from the right over the index array $\gamma$. 

\noindent \emph{ii}) We have already observed
that the action of $\cLFP$ mixes Dirichlet distributions with index
arrays in different \emph{bucket spaces}, corresponding to different
polynomial degrees. Since the symmetrization operator $\cS$ averages over
permutations, we need to clarify how it behaves with
respect to a shift of degree. This is needed in order to generate a
dictionary of reference formulae analogous to
eqs.~(\ref{eq:firstalg})--(\ref{eq:lastalg}). From a theoretical point
of view, the problem originates from the fact that the index
raising/lowering operators do not commute with permutations. Indeed,
they fulfill the relations 
\begin{align} 
\label{eq:opluscommut}
\oplus_\ell\cdot \sigma & = \sigma\cdot \oplus_{\sigma(\ell)}\\[1.0ex]
\ominus_\ell\cdot \sigma & = \sigma\cdot
\ominus_{\sigma(\ell)}\,,\qquad \ell=1,\ldots,Q-1\,,\qquad
\sigma\in S_{Q-1}\,.
\label{eq:ominuscommut}
\end{align}
As an example, we notice that the action of $\phi_k^m\partial_k^n$ on
$\cS\cdot\cD_\gamma$ breaks the permutational symmetry of the symmetrized
Dirichlet distribution and shifts its degree for $m\ne n$. If we then sum
over $k$, the symmetry is recovered, but in general the result cannot be 
written anymore as a permutational average over
$\Lambda(\gamma)$. With the same spirit by which we wrote down
eqs.~(\ref{eq:firstalg})--(\ref{eq:lastalg}), we consider some specific
cases. Two very simple ones are 
\begin{align}
  & \sum_{\ell=1}^{Q-1}\phi_\ell\,\cS\cdot\cD_\gamma(\bar\phi) =
  \frac{1}{|\Lambda(\gamma)|}\sum_{\eta\in\Lambda(\gamma)}|\bar\phi|\,\cD_\eta(\bar\phi)
  =
  \frac{1}{|\Lambda(\gamma)|}\sum_{\eta\in\Lambda(\gamma)}(s+|\bar\phi|-s)\,\cD_\eta(\bar\phi)
  \nonumber\\[1.0ex]
  & \hskip 0.2cm = s\,\cS\cdot\cD_\gamma(\bar\phi) - s\frac{\gamma_Q}{|\gamma|}\,\cS\cdot\cD_{\gamma_{Q^+}}(\bar\phi)\,,
\label{eq:firstanalytic}
\end{align}
and
\begin{align}
  & \sum_{\ell=1}^{Q-1}\phi_\ell
   \partial_\ell\,\cS\cdot\cD_\gamma(\bar\phi)  =
   \frac{1}{|\Lambda(\gamma)|}\sum_{\eta\in\Lambda(\gamma)}\sum_{\ell=1}^{Q-1}\left[\theta_{\eta_\ell,2}(\eta_\ell-1)\cD_\eta(\bar\phi)-\theta_{\eta_Q,2}\eta_\ell\cD_{\eta_{\ell^+Q^-}}(\bar\phi)\right]\nonumber\\[1.0ex]
   & \hskip 0.2cm =
   \left[\sum_{\ell=1}^{Q-1}\theta_{\gamma_\ell,2}(\gamma_\ell-1)+\theta_{\gamma_Q,2}(\gamma_Q-1)\right]\,\cS\cdot\cD_\gamma(\bar\phi) 
   -\theta_{\gamma_Q,2}(|\gamma|-1)\,\cS\cdot\cD_{\gamma_{Q^-}}(\bar\phi)\,.
\label{eq:secondanalytic}
\end{align}
As can be seen, in both cases the result is still a permutational
average over 
$\Lambda(\gamma)$. However, let us consider the action of
$\sum_{k=1}^{Q-1}\partial_k$ on $\cS\cdot\cD_\gamma$. From
eq.~(\ref{eq:delDgamma}), we have
\begin{align}
& \sum_{\ell=1}^{Q-1}\partial_\ell\,\cS\cdot\cD_\gamma(\bar\phi) =
  \frac{s^{-1}(|\gamma|-1)}{|\Lambda(\gamma)|}\sum_{\eta\in\Lambda(\gamma)} 
\sum_{\ell=1}^{Q-1}[\theta_{\eta_\ell,2}\cD_{\eta_{\ell^-}}(\bar\phi)
  - \theta_{\eta_Q,2}\cD_{\eta_{Q^-}}(\bar\phi)]\nonumber\\[2.0ex]
& \hskip 0.2cm = 
s^{-1}(|\gamma|-1)\left\{\left[\frac{1}{|\Lambda(\gamma)|}\sum_{\eta\in\Lambda(\gamma)}\sum_{\ell=1}^{Q-1}\theta_{\eta_\ell,2}\cD_{\eta_{\ell^-}}(\bar\phi)\right] 
- \theta_{\gamma_Q,2}(Q-1)\,\cS\cdot\cD_{\gamma_{Q^-}}(\bar\phi)\right\}\,.
\end{align}
In order to check that the sum in square brackets is
permutationally invariant, it is sufficient to make use of
eq.~(\ref{eq:ominuscommut}) and observe that
$\theta_{\eta_\ell,2}\cD_{\eta_{\ell^{-}}}(\sigma\cdot\bar\phi) =
\theta_{\gamma_{\sigma(\ell)},2}\cD_{\sigma\cdot\gamma_{\sigma(\ell)^{-}}}(\sigma\cdot\bar\phi) 
=
\theta_{\gamma_{\sigma(\ell)},2}\cD_{\gamma_{\sigma(\ell)^-}}(\bar\phi)$,
which upon summing over $\ell$ becomes manifestly invariant. Since
$|\slashed{\eta}_{\ell^-}|=n-1$ for $\eta\in\Lambda(\gamma)$, we can
write
\begin{equation} 
\frac{1}{|\Lambda(\gamma)|}\sum_{\eta\in\Lambda(\gamma)}\sum_{\ell=1}^{Q-1}\theta_{\eta_\ell,2}\cD_{\eta_{\ell^-}}(\bar\phi)
= \sum_{\eta\in\cL_{n-1}}\hat g_\eta\,\cS\cdot\cD_\eta(\bar\phi)\,,
\label{eq:deelsymexp}
\end{equation}
since we have already shown in eq.~(\ref{eq:Psym}) --- by an argument that
could be here repeated --- that a polynomial, which we know to be
symmetric, can be expanded as a linear combination of symmetrized
Dirichlet distributions (in this case the index arrays live on
$\cL_{n-1}$ due to the degree shift produced by the
differentiation). Determining the coefficients $\{\hat g_\eta\}$
analytically is non--trival and beyond the aims of this
paper. Nevertheless, by projecting eq.~(\ref{eq:deelsymexp}) onto the
symmetrized orthogonal polynomials and by making use of
eq.~(\ref{eq:symscalprod}), we obtain
\begin{equation}
\sum_{\eta\in\cL_{n-1}} (e_{\eta\alpha}f_\alpha)\hat g_\eta =
\frac{1}{|\Lambda(\gamma)|}\sum_{\eta\in\Lambda(\gamma)}\sum_{\ell=1}^{Q-1}\theta_{\eta_\ell,2}\langle 
\cS\cdot V_\alpha,\cD_{\eta_{\ell^-}}(\bar\phi)\rangle_\kappa\,,
\label{eq:firstnonanal}
\end{equation}
which can be numerically inverted. 

Along the same line, all the symmetric counterparts of
eqs.~(\ref{eq:firstalg})--(\ref{eq:lastalg}) can be worked
out. Analytic expressions such as
eqs.~(\ref{eq:firstanalytic})--(\ref{eq:secondanalytic}) 
effectively help save CPU time, while formulae requiring numerical
inversions such as eq.~(\ref{eq:firstnonanal}) are of no benefit. Such
cases require more sophisticated analyses, which we do not
attempt here. 

\section{Conclusions}

In this paper we have explored the possibility of representing
the solution of the Fokker--Planck equation for many--variable
steady--state birth--death systems as a linear combination of
Dirichlet distributions. This idea was first suggested in
\cite{palombi}, where a variant of the multi--state
voter model with zealots over a community--based network
\cite{Girvan} was studied. We have shown here that \emph{quasi}--optimal
coefficients for such a linear expansion can be generally obtained
from a variant of the Ritz--Galerkin method for partial differential
equations. As a test, we have applied the Dirichlet expansion
successfully to the binary and multi--state voter models with zealots
on a complete graph. Although Ritz--Galerkin techniques are widely
employed in engineering applications, no adaptation to systems defined
on the simplex has been ever considered in the literature, to the best
of our knowledge. 

We expect the domain of applicability of the method to go beyond that
of voter models and to extend to a \emph{positive--measure subset} of
statistical physics. Applications could include variants of SIS model
for epidemic spreading, naming games (a variant with committed agents
has been recently studied in \cite{Xie2}) and other complex
systems, only subject to the conditions that \emph{i}) a steady--state
distribution with positive variances exists and \emph{ii}) the drift
and diffusion coefficients of the Fokker--Planck equation are
polynomials. With regard to condition \emph{i}), our proposal could be
generalized by considering an expansion in Dirichlet  distributions
with linear coefficients depending on time, so as to allow for a
treatment of the time--dependent Fokker--Planck equation. This would
permit to describe the system while it relaxes to
equilibrium. Nevertheless, systems with consensus--like exit states,
which have 
attracted much attention in recent years, are anyway ruled out as
finite--degree polynomials can never approximate a Dirac delta
distribution. Concerning condition~\emph{ii}), cases where the drift
and diffusion coefficients are non--polynomial analytic functions
could be maybe faced by expanding them in power series to some finite
degree $m$ and by subsequently applying the Ritz-Galerkin method; this
would generate a sequence $(\cP_{n,m})$ of solutions, whose
convergence for $n,m\to\infty$ should be studied.  

We conclude by recalling that in this work we have provided no
theoretical arguments to show that the coerciveness condition of the
Lax--Milgram theorem is generally fulfilled by a Fokker--Planck
operator with polynomial diffusion matrix on the simplex. Without a
general proof, the applicability of the method has to be checked on a
case--by--case basis.  

\section*{Acknowledgments}

The computing resources used for our numerical study and the related technical support have been partly provided by the CRESCO/ENEA\-GRID High Performance Computing infrastructure and its staff \cite{cresco}. CRESCO ({\color{red}C}omputational  {\color{red} RES}earch centre on {\color{red} CO}mplex systems) is funded by ENEA and by Italian and European research programmes.

\begin{appendices}

\section{Dirichlet integrals}

\subsection{Normalization of $\cD_\alpha$}

It is worthwhile describing a simple technique to calculate polynomial
integrals on the simplex by means of a specific example, namely the
normalization coefficient of the Dirichlet distribution
$\cD_\alpha(\bar\phi) = Z_{\cD}(\alpha,s)^{-1}
\phi_1^{\slashed{\alpha}_1}\ldots
\phi_{Q-1}^{\slashed{\alpha}_{Q-1}}\left(s-|\bar\phi|\right)^{\slashed{\alpha}_Q}$. Specifically,
the integral that we aim at calculating is 
\begin{equation}
Z_\cD(\alpha,s) = \int_{\bar T_Q(s)}\rd
\bar\phi\ \phi_1^{\slashed{\alpha}_1}\ldots
\phi_{Q-1}^{\slashed{\alpha}_{Q-1}}\left(s-|\bar\phi|\right)^{\slashed{\alpha}_Q}\,. 
\end{equation}
By introducing a Dirac delta function, this integral can be brought to
the equivalent form 
\begin{equation}
Z_\cD(\alpha,s) = \int_0^s\rd \phi_1\ldots \int_0^s\rd \phi_{Q}
\ \phi^{\slashed{\alpha}}\delta\left(s-|\phi|\right)\,. 
\end{equation}
Moreover, owing to the Dirac delta function, all the upper integration
limits can be pushed to infinity without changing the integral. If we
replace the Dirac delta function by its Fourier representation 
\begin{equation}
\delta(z) = \frac{1}{2\pi}\int_{-\infty}^{+\infty}\rd \lambda \re^{-i\lambda z}\,,
\end{equation}
and rotate $\lambda\to i\lambda$, the integral turns into a complex
one, performed along the imaginary axis, \ie 
\begin{equation}
Z_\cD(\alpha,s) = \frac{1}{2\pi i}\int_{-i\infty}^{+i\infty}\rd \lambda\ \re^{\lambda s}\ \left(\prod_{k=1}^{Q} \int_0^{+\infty}\rd \phi_k\ \phi_k^{\alpha_k-1} \re^{-\lambda \phi_k}\right)\,,
\end{equation}
The inner integrals are Laplace transforms of monomials. They sum to 
\begin{equation}
\int_0^{+\infty}\rd \phi\ \phi^{\alpha-1} \re^{-\lambda \phi} = \frac{\Gamma(\alpha)}{\lambda^{\alpha}}\,.
\end{equation}
Hence it follows
\begin{equation}
Z_\cD(\alpha,s) = \dfrac{\prod_{k=1}^{Q}\Gamma(\alpha_k)}{2\pi
  i}\int_{-i\infty}^{+i\infty}\rd \lambda\ \frac{\re^{\lambda
    s}}{\lambda^{|\alpha|}}\  =
\dfrac{\prod_{k=1}^{Q}\Gamma(\alpha_k)}{\Gamma(|\alpha|)}
s^{|\alpha|-1}\,, 
\label{eq:wk}
\end{equation}
as a result of the Laplace antitransform of $\lambda^{-|\alpha|}$.

\subsection{Integrals of $V_\alpha$ and $U_\alpha$ on $\bar T_Q(s)$}

The integral of $V_\alpha$ on $\bar T_Q(s)$ follows trivially from
a term--by--term integration of its monomial expansion, namely
\begin{align}
Z_V(\alpha,s) & \equiv \int_{\bar T_Q(s)}\rd\bar\phi\,
V_\alpha(\bar\phi) \ = \sum_{\beta\le\alpha}v_{\alpha\beta}(\kappa)
\frac{\prod_{m=1}^{Q-1}\Gamma(\beta_m+1)}{\Gamma(|\beta|+Q)}\,.
\label{eq:Vint}
\end{align}
The integral of $U_\alpha$ on $\bar T_Q(s)$ can be similarly
calculated, provided we first represent it as a monomial sum. To this
aim, 
we just need to apply the standard binomial formula 
\begin{equation}
\partial_x^{k}[f(x)g(x)] = \sum_{\ell = 0}^k {k\choose
  l}[\partial_x^{k-\ell}f(x)][\partial_x^\ell g(x)]\,, 
\end{equation}
in sequence to the various factors of eq.~(\ref{eq:Ubasis}). We also
observe that, given $k,m\in\dN$ with $k\le m$, it holds 
\begin{align}
& \partial_x^k x^m = k!{m\choose k}x^{m-k}\,,\\[2.0ex]
& \partial_x^k (s-x)^{m} = (-1)^k k!{m\choose k}(s-x)^{m-k}\,.
\end{align}
Accordingly, we have
\begin{equation}
U_\alpha(\bar\phi) =
\frac{1}{\cD_\kappa(\bar\phi)}\partial_2^{\alpha_2}\ldots\partial_{Q-1}^{\alpha_{Q-1}}
P_1(\bar\phi)\,. 
\end{equation}
with
\begin{align}
P_1(\bar\phi) & = \phi_2^{\alpha_2+\kappa_2-1}\ldots
\phi_{Q-1}^{\alpha_{Q-1}+\kappa_{Q-1}-1}\partial_1^{\alpha_1}
\left\{\phi_1^{\alpha_1+\kappa_1-1}(s-|\bar\phi|)^{|\alpha|+\kappa_{Q}-1}\right\}
\nonumber\\[2.0ex]
& =
\alpha_1!\sum_{\eta_1=0}^{\alpha_1}(-1)^{\eta_1}{\alpha_1+\kappa_1-1\choose
  \alpha_1-\eta_1}{|\alpha|+\kappa_{Q}-1 \choose
  \eta_1}\nonumber\\[2.0ex] 
& \cdot\phi_1^{\eta_1+\kappa_1-1}\phi_2^{\alpha_2+\kappa_2-1}\ldots
\phi_{Q-1}^{\alpha_{Q-1}+\kappa_{Q-1}-1}(s-|\bar\phi|)^{|\alpha|-\eta_1+\kappa_{Q}-1}\,.  
\end{align}
Upon iterating the above calculation over all derivatives, we arrive at
\begin{align}
U_\alpha(\bar\phi) & =
\left[\prod_{m=1}^{Q-1}\Gamma(\alpha_m+1)\right]\sum_{\eta\le\alpha}(-1)^{|\eta|}\prod_{m=1}^{Q-1}{\alpha_m
  + 
  \kappa_m-1\choose \alpha_m-\beta_m}{|\alpha|-L_m+\kappa_{Q}-1\choose
  \eta_m}\nonumber\\[2.0ex] 
& \hskip 2.0cm \cdot
\phi_1^{\eta_1}\ldots\phi_{Q-1}^{\eta_{Q-1}}(s-|\phi|)^{|\alpha|-|\eta|}\,, 
\end{align}
with $L_1\equiv 0$ and $L_m \equiv \sum_{k=1}^{m-1}\eta_k$ for $m\ge
2$. Therefore, we have 
\begin{align}
& Z_U(\alpha,s) \equiv \int_{\bar T_Q(s)}\rd\bar\phi\,U_\alpha(\bar\phi)  =
  \frac{\prod_{m=1}^{Q-1}\Gamma(\alpha_m+1)}{\Gamma(|\alpha|+Q)} \nonumber\\[2.0ex]  
& \hskip 1.0cm  \cdot
  \sum_{\eta\le\alpha}(-1)^{|\eta|}\,\Gamma(|\alpha|-|\eta|+1)
  \prod_{m=1}^{Q-1}\Gamma(\eta_m+1){\alpha_m+\kappa_m-1\choose
    \alpha_m-\eta_m}{|\alpha|-L_m+\kappa_{Q}-1\choose \eta_m}\,. 
\end{align}

\end{appendices}

\bibliographystyle{hunsrt}
\bibliography{main}

\end{document}